\documentclass[a4paper,11pt]{article}
\usepackage{jcappub}
\usepackage{orcidlink}

\title{Generation of Imaging Air Cherenkov Telescope images using Diffusion Models}

\affiliation[a]{
  Erlangen Centre for Astroparticle Physics,
  Friedrich-Alexander-Universit\"at Erlangen-N\"urnberg\\
  Nikolaus-Fiebiger-Str. 2, 91058 Erlangen, Germany}
\affiliation[b]{Kobayashi-Maskawa Institute, Nagoya University, \\Aichi 464-8602, Japan}
\affiliation[c]{Department of Particle Physics and Astrophysics, Stanford University, \\Stanford, CA 94305, USA}
\affiliation[d]{Fundamental Physics Directorate, SLAC National Accelerator Laboratory, \\Menlo Park, CA 94025, USA}
\affiliation[e]{Department of Physics, University of California, \\Berkeley, CA 94720, USA}

\emailAdd{christian.elflein@fau.de}
\emailAdd{s.funk@fau.de}
\emailAdd{jonas.glombitza@fau.de}
\emailAdd{vmikuni@hepl.phys.nagoya-u.ac.jp}
\emailAdd{nachman@standford.edu}
\emailAdd{lark\_wang@berkeley.edu}

\author[a]{Christian Elflein\orcidlink{0009-0008-0854-8480},}
\author[a]{Stefan Funk\orcidlink{0000-0002-2012-0080},}
\author[a, 1]{Jonas Glombitza\orcidlink{0000-0001-9683-4568}\note{Work partly done at Physics Division, Lawrence Berkeley National Laboratory, Berkeley, CA 94720, USA.},}
\author[b]{Vinicius Mikuni\orcidlink{0000-0002-1579-2421},}
\author[c,d]{Benjamin Nachman\orcidlink{0000-0003-1024-0932},}
\author[e]{and Lark Wang}

\abstract{
Substantial amounts of air-shower simulations are needed to derive the instrument response for analyzing Imaging Air Cherenkov Telescope (IACT) data.
This process is both computationally intensive and requires repetition under varying observation conditions, due to detector aging, changes in the atmosphere, or the instrument hardware.
Generative models offer an efficient alternative, significantly accelerating simulations while compactly storing extensive simulation libraries, and providing a differentiable surrogate model of the instrument.
However, their applicability has so far been limited in gamma-ray astronomy, particularly for modeling hadronic showers that dominate the background and exhibit significant intrinsic fluctuations that are challenging to model. 

In this study, we present the first application of score-based diffusion models to generate monoscopic $\gamma$-ray and proton shower images with nearly 2,000 pixels and benchmark the performance against Wasserstein GANs using H.E.S.S. simulations.
We examine quality using both low-level parameters and well-established shower-shape observables, and take the first step towards analysis readiness by investigating state-of-the-art $\gamma$–hadron separation.
While GAN-based approaches can reproduce $\gamma$-ray showers with high fidelity, they fail to generate proton events of comparable quality, particularly in capturing intrinsic higher-order correlations, leading to a measurable degradation in analysis performance.
In contrast, score-based diffusion models achieve significantly superior quality for $\gamma$-ray and proton showers, accurately reproducing high-level correlations and generating events that are statistically indistinguishable from simulations at the analysis level.
These results establish diffusion-based models as the first analysis-ready surrogate model of a single IACT, opening new prospects for fast instrument response generation, detector optimization, and connected downstream tasks.
}

\begin{document}
\maketitle
\flushbottom

\section{Introduction}
Over the past two decades, $\gamma$-ray observatories, and particularly arrays of Imaging Air Cherenkov Telescopes (IACTs), have changed our understanding of the very-high-energy (VHE) $\gamma$-ray sky.
A key facility is the High Energy Stereoscopic System (H.E.S.S.) in Namibia, which studies cosmic $\gamma$ rays with energies in the GeV to TeV range~\cite{hess_crab} in the Southern hemisphere. IACTs image Cherenkov radiation emitted from secondary particles in extensive air showers induced by cosmic particles.
The recorded IACT images are then analyzed to extract information about the primary particle.
Traditionally, the so-called Hillas parameters~\cite{hillas1985cerenkov}, which describe the shape of the shower, are used to estimate the energy and reconstruct the arrival direction of the incoming primary particle, as well as to reject hadronic showers that outnumber $\gamma$-rays by $1$ to $10^3-10^4$.

\begin{figure}[t]
  \centering
  \includegraphics[scale=0.425]{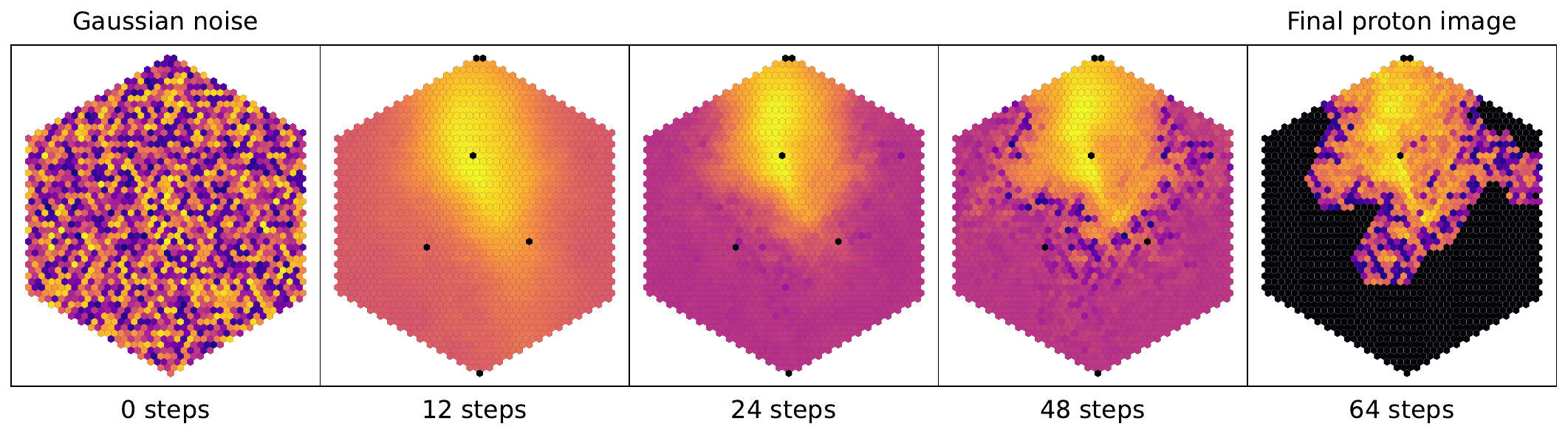}
  \caption{
  Various stages of the generation process of an IACT image starting from Gaussian noise using a score-based diffusion model.
  The particle causing the signal is a proton with an energy of 5.5 TeV.
  The black pixels denote zero-value pixels and brighter pixels denote higher values.}
  \label{fig:step_comparison}
\end{figure}

The analysis of $\gamma$-ray data requires a thorough understanding of the instrument response functions (IRFs), which are used to interpret observations by forward-folding theoretical models, e.g., by using the \texttt{Gammapy} software~\cite{Gammapy}.
Large-scale Monte Carlo (MC) simulations are required to precisely derive the IRFs.
Running these simulations, particularly for the hadronic background, is computationally demanding and time-intensive~\cite{heck_corsika_1998, simtel}, as classifiers reach rejection powers in the order of $10^{3}$ and higher.
Running millions of events, each requiring in the order of 30--60 seconds on one CPU core, generates substantial data, typically hundreds of gigabytes per observation period.
Furthermore, as the instrument ages, new simulations are needed to account for small variations in performance.

Recent advances in deep learning~\cite{lecun_deep_2015} provide novel perspectives on data analysis in physics\footnote{For a pedagogical introduction, see ref.~\cite{dlfpr}.} and high-energy-physics research~\cite{feickert2021livingreviewmachinelearning, Guest_2018}.
Particularly, generative models offer novel opportunities for simulating high-dimensional data~\cite{Hashemi_2024, calochallenge}.
In particle physics, Generative Adversarial Networks (GANs)~\cite{GanPhys2, Paganini:2017dwg, Erdmann:2018jxd, Aad_2022, mazurek2025machinelearninglhcbsimulation}, normalizing flows~\cite{Krause:2021ilc}, and diffusion models~\cite{mikuni:caloscore} have been used to construct surrogates of the instrument and accelerate simulations by five to six orders of magnitude.
Deriving these surrogate models can be used to adapt simulations to slightly different conditions~\cite{wang2018transferring, Diefenbacher_2020} and refine simulations to match measured data~\cite{Erdmann:2018kuh}.
Furthermore, these differentiable surrogates unlock a broad spectrum of downstream tasks, including instrument optimization~\cite{dorigo2022endtoendoptimizationparticlephysics}, event reconstruction~\cite{Du_2025}, anomaly detection~\cite{Krause_2024_anomaly}, unfolding~\cite{Huetsch_2025}, and the modeling of non-perturbative processes, such as hadronization~\cite{ilten2022modelinghadronizationusingmachine, Ghosh_2022_hadronization}, where first-principles descriptions are currently unavailable.
Addressing the computational needs of future observatories like the Cherenkov Telescope Array Observatory (CTAO)~\cite{CTA}, recent studies have explored generative models for fast and memory-efficient generation of high-level IACT parameters~\cite{tactic} or complete IACT images using GANs~\cite{taiga_1, taiga_2, veritas_wgan, wgan_veritas}, to obtain surrogates of the full-scale simulations.

Also, in recent years progress has been made in deep-learning-based reconstructions for $\gamma$-ray observatories~\cite{Shilon_2019, ct_learn, Brill_2019, Jacquemont_2021, Spencer_2021, Glombitza_2023, WCD4PMTs, schwefer2024hybridapproacheventreconstruction, Glombitza_2025} and cosmic-ray observatories~\cite{thepierreaugercollaboration2024inference, ERDMANN201846,pao_muon_dnn, xmax_wcd}.
Early studies have already produced convincing results for generating monoscopic images of $\gamma$-ray showers~\cite{Elflein_2024, diffusion_astro_ml4jets}.
However, generating detailed images of Imaging Atmospheric Cherenkov Telescopes (IACTs) with comprehensive fidelity remains a challenge, particularly for hadrons, for which large shower-to-shower fluctuations in the shower development exist.
These fluctuations are characterized by distinctive subshowers, manifesting themselves as hadronic substructure in the Cherenkov images, making them more irregular and patchy --- distinctly different from the relatively homogeneous and elliptical shapes of $\gamma$-ray images.
Consequently, while current generative models can reproduce $\gamma$-ray image features, their limited ability to capture these hadronic fluctuations prevents the models from being analysis-ready.

Inspired by the application in particle physics calorimeters~\cite{mikuni:caloscore, Mikuni_2024}, we use for the first time state-of-the-art score-based diffusion models (SBDM)~\cite{Song2021ScoreBasedGM} in astroparticle physics.
Using simulations from the H.E.S.S. CT5 telescope with its FlashCam design~\cite{puehlhofer2021science}, the camera of the CTAO MSTs on the southern site, we study the application of diffusion models to the generation of IACT images.
A simple illustration of the generation process using the SBDM is shown in figure~\ref{fig:step_comparison} using an IACT image from a proton-induced air shower.
We demonstrate that diffusion models provide accurate generation of both $\gamma$-ray and proton images, outperforming previous approaches based on generative adversarial networks (GANs), which show good performance in generating $\gamma$-ray images, providing a meaningful baseline.
Finally, we investigate the readiness for analysis of both approaches by applying state-of-the-art $\gamma$-hadron separators to the events generated by the WGAN and SBDM.
Since stereoscopic integration is an ongoing challenge in deep learning~\cite{Shilon_2019, Brill_2019, Jacquemont_2021, Spencer_2021, Glombitza_2023, Warnhofer_2024}, we limit ourselves in this work to monoscopic, i.e., single-telescope images.

\section{Generation of IACT images using deep learning}
In recent years, particularly generative adversarial networks (GANs), have gained significant attention for their impressive results in generating natural images and have been applied to many different challenges in particle and astroparticle physics~\cite{GanPhys1, GanPhys2, Paganini:2017dwg, Paganini:2017dwg, Erdmann:2018jxd, Erdmann:2018kuh, veritas_wgan, tactic, Aad_2022, taiga_1, taiga_2, Yang2020PhysicsInformedGA}.
Most recently, diffusion models~\cite{DBLP:journals/corr/abs-2010-02502, ho2020denoisingdiffusionprobabilisticmodels} gained popularity by significantly improving over GANs in both image quality and fidelity.
In this section, we first introduce Wasserstein GANs, the current state-of-the-art technique in IACT image generation, and the baseline in this work.
Subsequently, we discuss diffusion models which have not yet been applied to astroparticle physics and will be used in this analysis to challenge the current state of the art.

\subsection{Wasserstein Generative Adversarial Networks}
\label{sec:wgan}
The Generative adversarial networks (GANs) proposed in ref.~\cite{Goodfellow:2014upx} comprise two networks, a generator \(G\) and a discriminator \(D\), which are trained adversarially.
The generator in the framework is trained to generate realistic-looking samples $x_r$ from a distribution of real samples $p_r$, using the discriminator's feedback.
While the generator maps noise \(z\) from a simple distribution \(p_z\) to realistic and high-dimensional samples \(x_{\theta} = G(z)\), the discriminator is trained to discriminate real \(x_r\) from generated samples \(x_\theta\).
The consecutive training of the generator and discriminator networks involves the following zero-sum loss:
\begin{equation}
\begin{aligned}
\mathcal{L}_{\text{GAN}} = \min_G \max_D \; \mathbb{E}_{x_r \sim p_r}\left[\log(D(x_r))\right] + \mathbb{E}_{z\sim p_z}\left[\log(1-D(G(z)))\right].
\end{aligned}
\end{equation}

The discriminator loss is the binary cross-entropy objective, and therefore is trained to classify between real and generated samples.
In contrast, the generator minimizes the loss of the discriminator, i.e., it is trained to generate convincing, realistic-looking samples that fool the discriminator.
Throughout training, the adaptive parameters of each network are fixed alternately in consecutive training steps to ensure meaningful updates.

In this work, we make use of Wasserstein GANs, an improved version of GANs that significantly improves sample quality, diversity, and training stability~\cite{arjovsky2017principled, arjovsky2017wasserstein}.
Instead of fooling the cross-entropy-trained discriminator, in the WGAN approach, the discriminator is trained to approximate the Wasserstein distance between the distribution of generated and real samples. The Wasserstein distance provides a similarity metric for high-dimensional distributions and, in particular, provides reasonable estimates for disjoint distributions.
In turn, the generator is trained to minimize the Wasserstein distance and thus improves the quality of the generated samples by making the distributions more similar.
The min-max objective of the generator $G$ and the discriminator $D$ in the WGAN training reads:
\begin{equation}
\begin{aligned}
\mathcal{L}_{\text{WGAN-GP}} =  \min_G \max_D \; \mathbb{E}_{x_r\sim p_r}[D(x_r)] - \mathbb{E}_{z\sim p_z}[D(G(z))] + \lambda\cdot\mathcal{L}_\mathrm{GP},
\end{aligned}
\end{equation}
where $\lambda$ is a hyperparameter to scale the gradient penalty~\cite{gulrajani2017improved} loss:
\begin{equation}
    \mathbb{E}_{x_u\sim p_u}[(\left\lVert \nabla D(x_u)\right\rVert _2 - 1)^2]
\end{equation}
enabling the discriminator to approximate the Wasserstein distance, by regularizing the gradients of the discriminator to be 1 on mixture samples $x_u = \epsilon \cdot x_r + (1-\epsilon) \cdot x_{\theta},\; \text{where} \; \epsilon \sim \mathcal{U}[0,1]$.
To condition the generated properties on specific kinematics, such as for example the energy $E$, the generation process is adapted by $G(z) \rightarrow G(z, E)$, and extending the generator loss by a conditioning reconstruction term
\begin{equation}
     \mathcal{L}'_\mathrm{WGAN-GP} = \min_G \max_D
     \,\mathcal{L}_\mathrm{WGAN-GP} -  \mathbb{E}_{z\sim p_z}\left(\hat{E}(G(z)) - E\right)^2,
\end{equation}
which involves another neural network $\hat{E}$ trained to reconstruct the given quantity using the mean squared error as a loss.

\paragraph{Baseline design}
The baseline in this work is the WGAN architecture introduced in ref.~\cite{Elflein_2024} for the generation of $\gamma$-ray showers, which uses two conditioning networks on the impact point and the energy of the particle.
All networks make use of residual connections~\cite{he2016deep} of convolutional layers~\cite{dlfpr}.

\subsection{Diffusion Models}
Diffusion generative models have shown state-of-the-art quality in image synthesis and promising performance as fast surrogate models for expensive physics simulations.
In particular, diffusion models used to reproduce the detector response of calorimeters in collider experiments such as \textsc{CaloScore}~\cite{mikuni:caloscore} have shown improved fidelity compared to previous deep-learning approaches.
Another advantage of diffusion models is the simplicity of the training procedure compared to alternative strategies.
Diffusion models are trained to predict how much noise is present in a dataset as a function of a time parameter $t$.\footnote{Although referred to as time, $t$ is not related to the real time of the particles that interact with the detector.}
The diffusion process is created such that at $t=1$ the dataset is dominated by a noise distribution, often chosen to be a standard normal distribution.
As time decreases and approaches $t=0$, the distribution gradually evolves from complete noise towards the desired dataset, in our case consisting of images with Cherenkov light emitted by $\gamma$-ray and proton showers and detected by the telescope.
We can model this evolution by defining the time-dependent dataset $\mathbf{x_t} = \mathbf{x(t)}$ as:
\begin{equation}
    \mathbf{x_t} = \alpha(t)\mathbf{x} + \sigma(t)\mathbf{\epsilon},
    \label{eq:perturb}
\end{equation}
where $\mathbf{\epsilon}\sim\mathcal{N}(0,1)$ and the time-dependent functions $\alpha$ and $\sigma$ determine the mixture between clean examples $\mathbf{x}$ and noise $\mathbf{\epsilon}$.

During training, we take our initial examples $\mathbf{x}$ and randomly sample the parameter $t$ from a uniform distribution.
Next, we perturb the data according to~\ref{eq:perturb}, creating the input dataset that is passed to a neural network that takes as input both the perturbed data and the time parameter.
The goal of this network is to predict the amount of noise $\mathbf{\epsilon}$ to be injected into  the clean examples.
Variants of this training target were proposed to improve the stability of the training and to improve the quality of the datasets generated from the diffusion process.
In particular, we use the same implementation from $\textsc{CaloScore}$ where instead of predicting the noise $\mathbf{\epsilon}$ we predict $\mathbf{v}_t \equiv \alpha_t\mathbf{\epsilon}-\sigma_t\mathbf{x}$.
This velocity term is a linear combination of both noise and clean datasets and was shown to improve the generation quality of new samples. The loss to be minimized is then the square difference between the network prediction and the true velocity:

\begin{equation}
    \mathcal{L} = \mathbb{E}_{\bf{x}_t,t} \left\| \mathbf{v}_t - \mathbf{v}_{\theta}(\bf{x}_t,t)\right\|^2.
    \label{eq:loss_v2}
\end{equation}

\begin{figure}[t!]
  \centering
  \includegraphics[scale=.9]{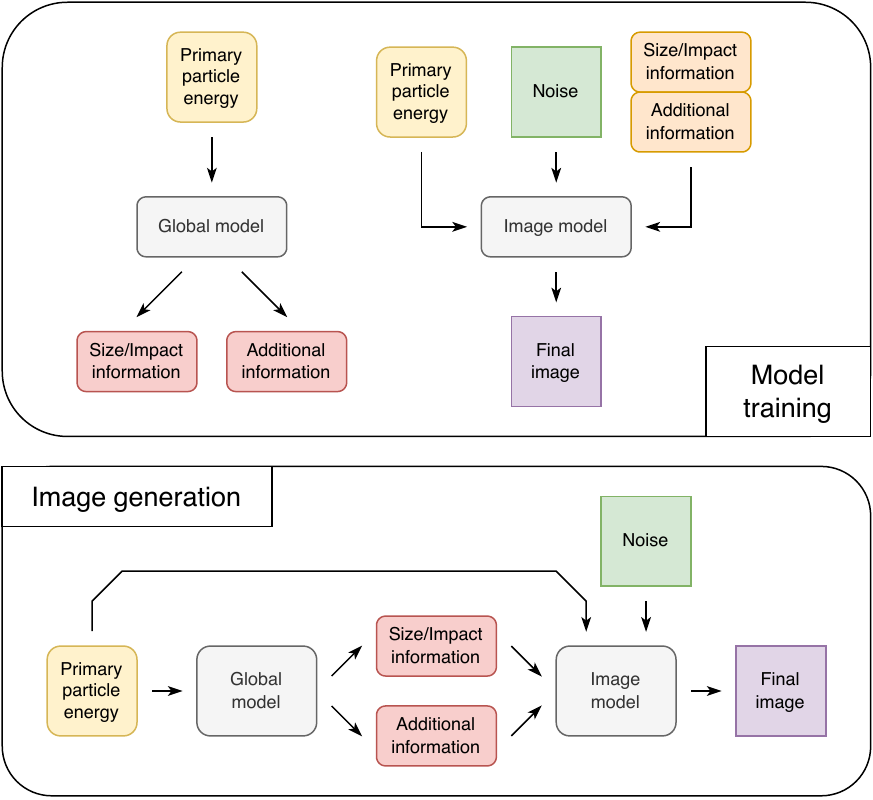}
  \caption{
  Sketch of the model training and image generation process of our diffusion model framework.
  During the model training, the global model learns to predict physical properties characterizing the images using only the primary particle energy as input.
  The image model learns to generate images from noise, energy, and the physical properties.
  During the image generation process the models are used together.
  The image model takes the information that the global model predicts alongside the input energy and noise to generate the final IACT image.
  }
  \label{fig:sketch_diffusion_model}
\end{figure}

In our implementation, we use a cosine parameterization for the perturbation parameters with $\alpha_t =\cos(0.5\pi t)$ and $\sigma_t=\sin(0.5\pi t)$.
In this parameterization, $\alpha^2 + \sigma^2 = 1$ for all times and the perturbed data slowly moves from clean datasets at $t=0$ to complete noise when $t=1$.
In order to generate new images, we use the approximate velocity learned by the neural network to generate samples by solving an ordinary differential equation with the DDIM solver~\cite{DBLP:journals/corr/abs-2010-02502}.
During sampling, we can control how many steps we allow the solver to use.
More steps result in slower generation time but in better generation quality.

To generate the images, we use the same network structure as used in \textsc{CaloScore}, consisting of a U-Net~\cite{ronneberger2015u} model, able to encode the information of the images in different resolution scales, combined with attention layers~\cite{vaswani2023attention} in the lower dimensions, so correlations in the embedding space of the network are properly learned through the attention mechanism.
As shown in \textsc{CaloScore}, we can further improve generation quality by breaking down the model into two components: the global model and the image model.
Our framework, consisting of the two models, is illustrated in figure~\ref{fig:sketch_diffusion_model}.
The image model learns to generate normalized images, where the sum of all energy deposits, i.e., the image size, is normalized to unity for all input images.
As input, it takes the primary particle energy, noise, and physical properties to condition the image generation.
To recover the image size, we train a second diffusion model based on the \textsc{ResNet}~\cite{he2016deep} architecture.
This model is trained to predict the image size, impact coordinates, and additional information\footnote{See appendix~\ref{appendix} for details on additional information as well as post-processing.} given the primary particle energy as input.
Image generation is then carried out using both models.
We input a particle energy into the global model, which predicts the size, impact, and additional information.
In addition to noise and energy, these outputs are subsequently used to generate images with the image model, conditioned not only on the time but also on the physical properties.

\section{Experimental setup}
In this work, we use simulations of the H.E.S.S.\ array, comprising the four original Cherenkov telescopes (CT1-4), arranged in a square of 120\,m side length, and a larger telescope (CT5) placed in the center.
H.E.S.S.\ features various detector configurations, characterized by the telescopes participating in the measurement~\cite{mono_mode}.
The mode used for our simulations is called \textit{mono} as it only utilizes the large CT5 telescope similar to previous work~\cite{Elflein_2024, diffusion_astro_ml4jets}.
The camera, which records the IACT images, uses the FlashCam design~\cite{Bi_2021} consisting of 1,758 photomultiplier tubes (PMTs), representing the pixels of the images.
Three pixels at the top and bottom of the camera are not used in measurements and were masked to zero in our data. Another set of three pixels around the inner part of the camera are not equipped with PMTs as they contain struts needed to mount the camera.
Although here the algorithm is applied to images observed by CT5, the concept can readily be used for the generation of images from other cameras with regular pixel alignment.

\subsection{Reference dataset}

MC simulations of both datasets were carried out with CORSIKA (COsmic Ray Simulation for KAscade)~\cite{heck_corsika_1998} to simulate the development of air-showers in the atmosphere and \texttt{sim\_telarray}~\cite{simtel} for the instrument response.
Hadronic interaction models for the air-shower development are EGS4~\cite{nelson1990egs4} and QGSJET~\cite{Ostapchenko_2014}.
The various configuration parameters are identical for the simulation of the datasets for the different models. The energies of the simulated events range from $10^{-1.5}$\,TeV to $10^{2}$\,TeV following an $E^{-2}$ spectrum. The zenith angle is set to $20^{\circ}$ and the opening angle of the telescopes is $5^{\circ}$, covering the field-of-view of CT5. 

Once the camera data is simulated, it undergoes a calibration process, which includes two main steps. First, the initial analog-to-digital converter (ADC) counts are converted into units of photoelectrons (p.e.).  Afterwards, the background data in the images is removed using the so-called tail-cuts cleaning~\cite{Malyshev_2023}, which is a noise reduction technique commonly used in $\gamma$-ray astronomy~\cite{Celic_2025, Unbehaun_2025}.
This method uses two threshold values to keep as much signal as possible and to remove most of the night sky background.
The extended 4/7 cleaning applied to our images keeps all pixel values higher than 4 p.e., under the condition that they have a neighboring pixel with a value higher than 7 p.e..
Additionally it keeps the next four rows, including negative pixels, around the cleaned signal.
This loosens the transition from the removed background to the signal in our images and results in better modeling of the signal edges.
The final images, which we apply our algorithm to, have an integrated signal above roughly 140\,p.e..
About 1.53 million $\gamma$-images and 630\,k proton images are used for this study.
15\,\% of each dataset is used for testing, while the remaining dataset is used for training and validation of the models.
The lower number of events in the proton datasets showed a negligible effect on the final performance of the trained models.
We further investigated pretraining to improve image quality, but found no improvement.

\subsection{Training strategy and generation time}
Both diffusion models, one for $\gamma$-ray images and one for proton images, are trained under the same conditions, with the only difference being the training time, as roughly twice as many $\gamma$-images as proton images are used.
The models are trained for 1500 epochs on 8 NVIDIA A100-SXM4-80GB, which took about 20\,h for the proton model and 44\,h for the $\gamma$-ray model.
The batch size is set to 4096 and the Exponential Moving Average (EMA) to 0.9999.
We use the Lion optimizer~\cite{chen2023symbolicdiscoveryoptimizationalgorithms} with Cosine decay with a warm-up learning rate of $10^{-5}$, which increases $10^{-4}$ after 10\,\% of the training steps, followed by a decrease of the learning rate to about $10^{-6}$ at the end of the training.
Furthermore, the optimizer uses the following parameters: $\beta_1 = 0.95$, $\beta_2 = 0.99$ and a weight decay of $0.1$.
In the generation process, the number of steps is chosen as 512 for the global model and 64 for the image model, striking a balance between performance and generation time.
The latter is chosen such that the generation time of the images is significantly reduced while maintaining an almost similar generation accuracy as for 512 steps.

The time required to generate 100\,k $\gamma$-ray and proton IACT images from simulations and generative models is shown in table~\ref{tab:gen_speed}, comparing GPU and CPU times.
As expected, the WGAN generates images significantly faster than the SBDM, since its prediction is performed in a single step, whereas the latter runs multiple steps iteratively during the denoising process.
Although the generation time of the WGAN and SBDM is similar for the $\gamma$-ray and proton images, this is not the case for the simulations, as the higher complexity in the hadronic interactions in proton air showers results in longer simulation times.

The WGAN generates images nearly 4000 times faster than MC simulations on a CPU.
With a GPU, it can produce 100,000 images in under three seconds, achieving a speed-up exceeding one million.
In contrast, the SBDM is slower, but still outpaces MC simulations. On a CPU, it shows a speed-up of about four to seven for $\gamma$-ray and proton images. When using a GPU, speed-ups of 2000 to 3000 can be reached.
It is important to note that this speed-up can be further improved as shown in ref.~\cite{Mikuni_2024} using progressive distillation~\cite{progressive_distillation}. 
Thus, the 64 diffusion steps used in this work can be reduced to four or even just a single step, resulting in another speed-up of one to two orders of magnitude.
Future work is planned to study this in detail as a reduction in diffusion steps is typically connected to a loss of generation quality.

\begin{table}[t!]
\centering
\begin{centering}
\begin{tabular}{ c | c | c | c | c }
\hline
& \multicolumn{2}{|c|}{$\gamma$-ray images} & \multicolumn{2}{|c}{Proton images} \\
\hline
Method & Time & \textbf{Speed-up} & Time & \textbf{Speed-up} \\
\hline
MC (1 CPU core) & 860\,h  & -- & 1320\,h  & -- \\
WGAN (1 CPU core) & 14\,min & \textbf{x\,3700} & 14\,min & \textbf{x\,5650}  \\
SBDM (1 CPU core) & 200\,h & \textbf{x\,4.3} & 200\,h & \textbf{x\,6.6} \\
\hline
WGAN (1 GPU) & 2.6\,s & $\boldsymbol{\textbf{x}\,1.2 \cdot 10^6}$ & 2.6\,s &  $\boldsymbol{\textbf{x}\,1.8 \cdot 10^6}$\\
SBDM (1 GPU) & 27\,min & \textbf{x\,1900} & 27\,min & \textbf{x\,2900} \\
\hline
\end{tabular}
\caption{
Comparison of computational times required for obtaining 100,000 IACT images from different generation methods. The CPU for the MC simulation is an Intel Xeon E3-1240 v6, while the deep-learning methods use an AMD EPYC 7713 Milan and a NVIDIA A100-SXM4-80GB as CPU and GPU.
}
\label{tab:gen_speed}
\end{centering}
\end{table}

\subsection{Benchmark metrics}
\label{sec:benchmark}

\begin{figure}[t!]
  \centering
  \includegraphics[scale=0.515]{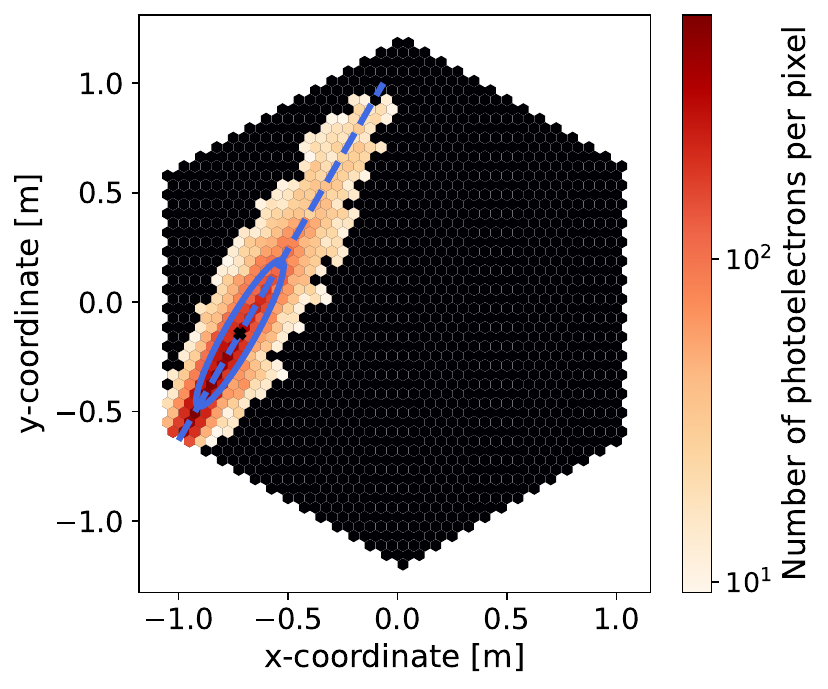}
  \hspace{0.25cm}
  \includegraphics[scale=0.515]{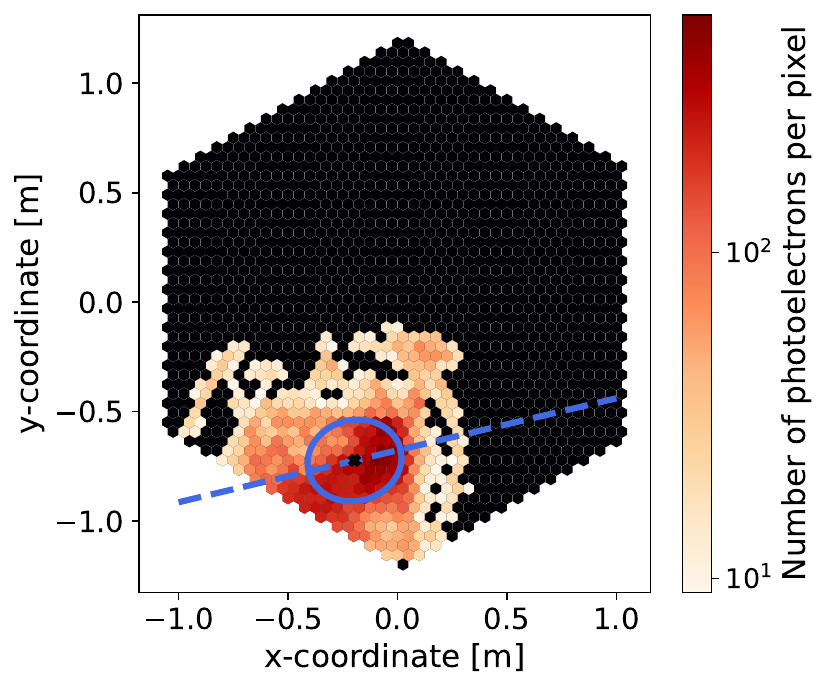}
  \caption{IACT $\gamma$-ray (left) and proton (right) image with applied tail-cuts cleaning using thresholds of 9\,p.e. and 16\,p.e..
  Included in both images is the reconstructed Hillas ellipse, its major axis, and center, respectively.
  The black pixels denote zero-value pixels.}
  \label{fig:hillas_ellipse}
\end{figure}

To properly verify that the deep learning models can generate realistic images, detailed analyses of the $\gamma$-ray and proton datasets have to be carried out. 
We begin the analysis by visually inspecting our generated and simulated images.
This is followed by an investigation of low-level observables that quantitatively describe the signal in the image.
Afterwards, the night sky background is removed using cleaning with thresholds of 9\,p.e. and 16\,p.e., which is the standard for H.E.S.S. analyses of CT5 images~\cite{Celic_2025}.
Then, the MC-simulated data and the data generated by the deep learning models are compared using the so-called ``Hillas parameters'' first introduced in ref.~\cite{hillas1985cerenkov}.
They are commonly used to characterize IACT images, separate $\gamma$-ray-induced images from cosmic-ray-induced images, and perform event reconstruction~\cite{Malyshev_2023}.
We used the open-source tool ctapipe~\cite{ctapipe-icrc-2023} (v0.21.2~\cite{karl_kosack_2024_12571953}) to perform the cleaning and reconstruction.
In figure~\ref{fig:hillas_ellipse}, a $\gamma$-ray image and a proton image are shown along with their moments, i.e., the COG, and the Hillas width and length.
The Hillas parameters we consider are the Hillas size, length, and width, the radial and polar coordinates, the rotation angle, the kurtosis, and the skewness.
The image intensity, often referred to as \textit{Hillas size}, is the sum of the signal after cleaning.
As the size is a measure of detected light in the camera, information about the primary particle energy and the distance to the shower core can be derived from this parameter.
For example, a large Hillas size corresponds to either a high-energy primary particle or a shower core close to the telescope.
The \textit{Hillas length} and \textit{Hillas width} is given as the RMS spread of the Cherenkov signal along the major and minor axis of the elliptical signal, respectively.
While the major axis is used to gain information about the source position, the minor axis gives information about the lateral spread of the air shower, which is important for background rejection.
The \textit{radial coordinate}, which is often referred to as local distance, is the distance of the center of gravity of the image signal to the center of the camera.
The angle between the ellipse center and the \textit{x}-axis of the camera is defined as the \textit{polar coordinate}.
Together, these two parameters return the location of the elliptical signal on the camera, providing information on the air-shower direction.
The \textit{rotation angle} is given as the angle between the major axis of the ellipse and the main axis of the camera.

In this study, we further investigate higher-order Hillas parameters that have not been considered in the evaluation of generated images before.
As a measure of the asymmetry of the Cherenkov signal along the major axis, we used the \textit{skewness} of the Hillas ellipse.
The sign of the skewness provides information about the direction of the asymmetry, its magnitude is proportional to the extent of the asymmetry, and in our analysis, we exclusively focus on the latter.
Lastly, we use the \textit{kurtosis} of the light distribution along the major axis as a measure of the peakedness and tail structure of the Cherenkov image, which is not captured in detail in the lower-order Hillas parameters.

To get an even deeper understanding of the Hillas parameters of the generated datasets, we investigate their correlations both visually through KDE contour plots and quantitatively using three different metrics.
For the quantitative approach, we first determine the Pearson correlations between the different sets of parameters resulting in the matrices $C_{\text{MC}}$, $C_{\text{WGAN}}$ and $C_{\text{SBDM}}$.
This is followed by the calculation of the \textit{Frobenius distance}, i.e., the application of the Frobenius norm $\|\cdot\|_{\mathrm{F}}$ to the difference of the correlation matrix of the simulated and generated dataset yielding:
\begin{equation}
\|C_{\text{MC}} - C_{\text{WGAN/SBDM}}\|_F = \sqrt{\sum_{i,j} (C_{\text{MC, }ij} - C_{\text{WGAN/SBDM, }ij})^2}.
\end{equation}
This metric shows how close the two matrices are to each other, with 0 representing identical matrices.
Next, we examine the \textit{correlation matrix similarity} by deriving the Pearson correlation between the lower triangular elements of $C_{\text{MC}}$ and $C_{\text{WGAN}}/C_{\text{SBDM}}$.
As the matrix is symmetric by design and the diagonal is 1, we exclude the upper triangular and diagonal elements.
The closer this final value is to 1, the more similar are the structures of the correlation matrices.
The uncertainties for both of those metrics are estimated using a bootstrapping approach with 1000 iterations.
Lastly, we use a simple two-class classification using a random forest with 200 trees to test the separation between simulated and generated correlations.
The metric investigated here is the classification \textit{accuracy}, which is 0.5 in the optimal case, indicating random guessing of the sample class.
The uncertainties are here estimated using k-fold cross-validation with $k=20$.

\section{Analysis of generated $\gamma$-ray images}
\label{sec:gamma}
Having introduced the deep-learning methods for image generation and the experimental setup, the analysis of the generated images is carried out.
In this section, the $\gamma$-ray images generated by the WGAN and SBDM are investigated, followed by an analysis of the proton images in the subsequent section~\ref{sec:proton}.
The aim of these analyses is to verify that deep-learning models can generate realistic IACT images and to benchmark their quality. After a minimal post-processing\footnote{See appendix~\ref{appendix} for details on the post-processing.} is carried out, about 227\,k images from the MC, WGAN, and SBDM dataset remain for this analysis.
Before examining any image parameters, the images need to be inspected for visual quality.
This is done in section~\ref{sec:gamma_visual} by examining handpicked images that portray the typical air-shower characteristics.
Next, in section~\ref{sec:gamma_low_level} and section~\ref{sec:gamma_high_level}, the low-level observables and high-level image Hillas parameters are investigated for a quantitative comparison.
As the final part of the analysis, the correlations between the high-level parameters are examined in section~\ref{sec:gamma_correlations} to demonstrate that even the most complex relations can be reproduced.

\subsection{Visual inspection}
\label{sec:gamma_visual}

\begin{figure}[t!]
  \centering
  \includegraphics[scale=0.37]{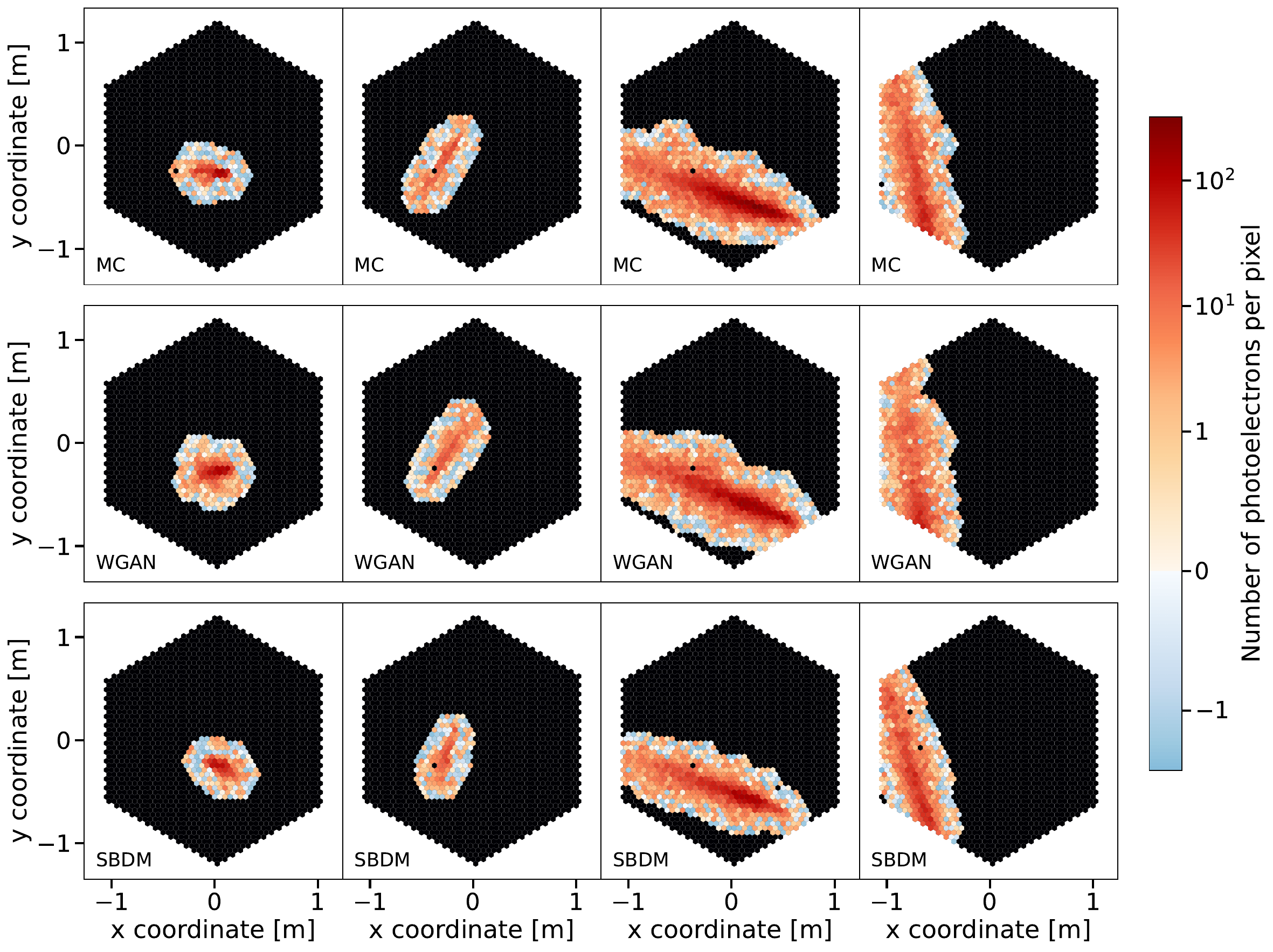}
  \caption{Comparison of four $\gamma$-ray IACT images obtained from MC simulations (top), the WGAN (middle), and the SBDM (bottom).
  The simulated images are hand-picked to show various characteristics of air showers induced by $\gamma$-rays.
  The WGAN and SBDM images are the next neighbors of the simulated images in the mean-squared-error pixel space.
  Note that a 4/7 tail-cuts cleaning with the extension of four rows is applied, and black pixels denote pixels that do not contain any signal.}
  \label{fig:four_gamma}
\end{figure}

Inspection of the visual quality of the generated images is an important step in this analysis.
Figure~\ref{fig:four_gamma} shows images from the three $\gamma$-ray datasets.
In the first row, four handpicked IACT images from the simulations are shown with typical signal shapes that one would expect from $\gamma$-ray showers. The second and third rows contain four images from the WGAN and SBDM datasets.
These images are obtained by calculating the Mean Squared Error (MSE) between the chosen MC images and the generated datasets and selecting the nearest neighbor.
In the first column, circular signals are shown, which are mostly related to air showers coming in at a low zenith angle.
The typical elliptical shape that we expect from IACT $\gamma$-ray images can be seen in the second and third columns.
The elongation of the signal is caused by imaging the longitudinal shower development in the atmosphere.
Furthermore, the distribution of the photoelectrons along the shower axis from low to high is also evident in all cases.
Lastly, in the fourth column, images are shown that are spatially truncated by the camera edge.
These showers are very frequent in any given dataset, and therefore it is especially important to examine this type of image to check how deep-learning models handle the camera edges and whether they correctly simulate the portion of the signal contained in the image.
Analysis of the three images indicates that generating truncated images does not seem to be a challenge for the models, as there are no irregularities at the edges and the remaining signal appears as expected.

\begin{figure}[t!]
  \centering
  \includegraphics[scale=0.35]{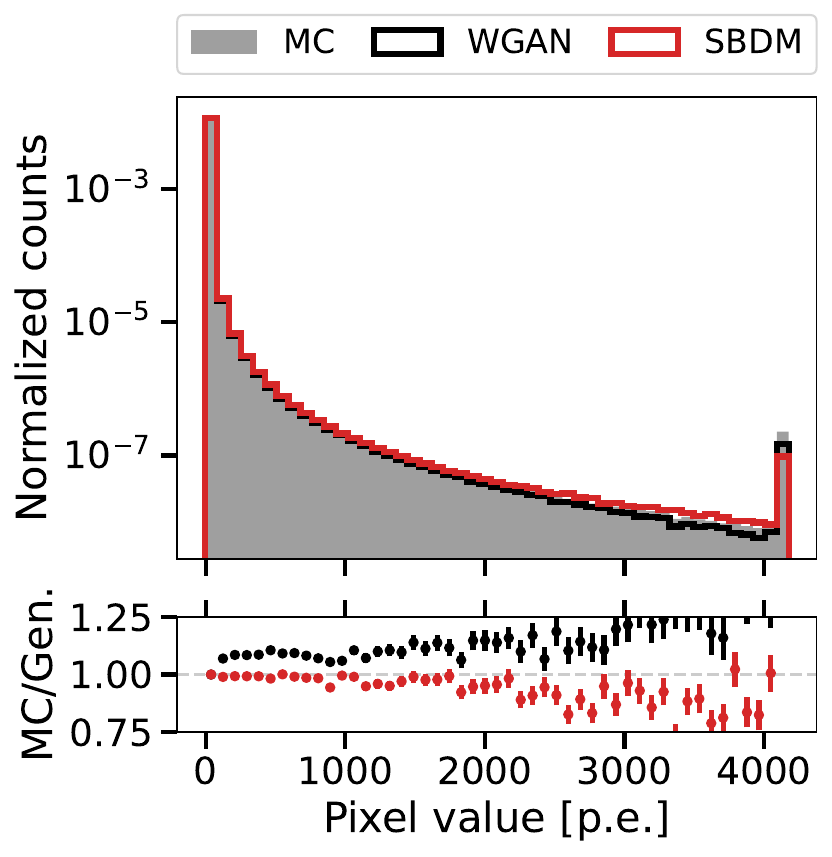}
  \includegraphics[scale=0.35]{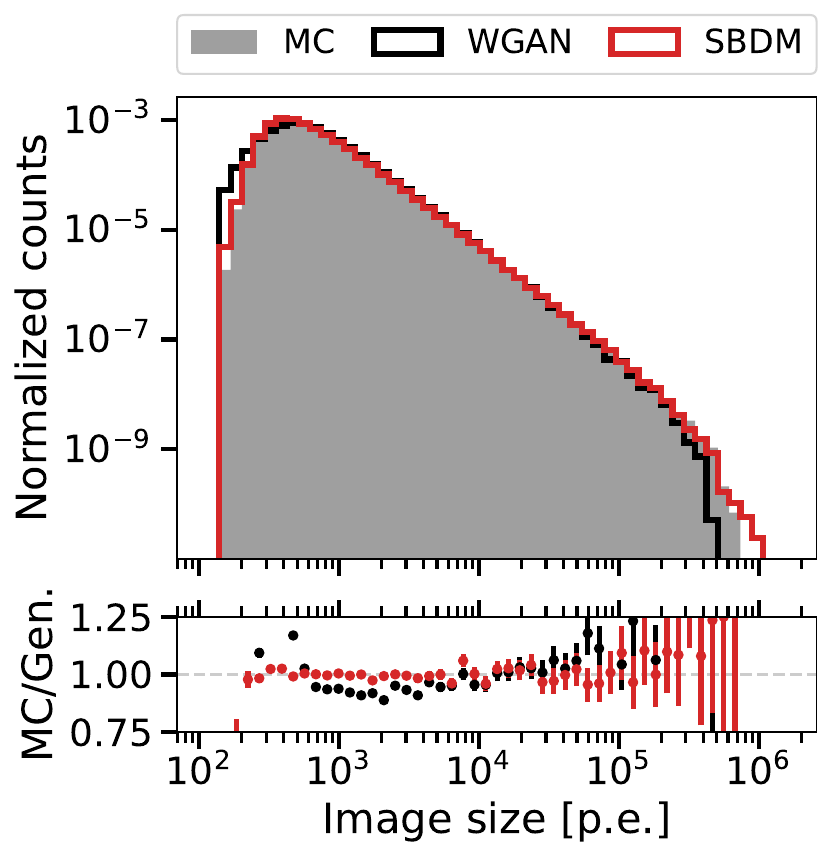}
  \includegraphics[scale=0.35]{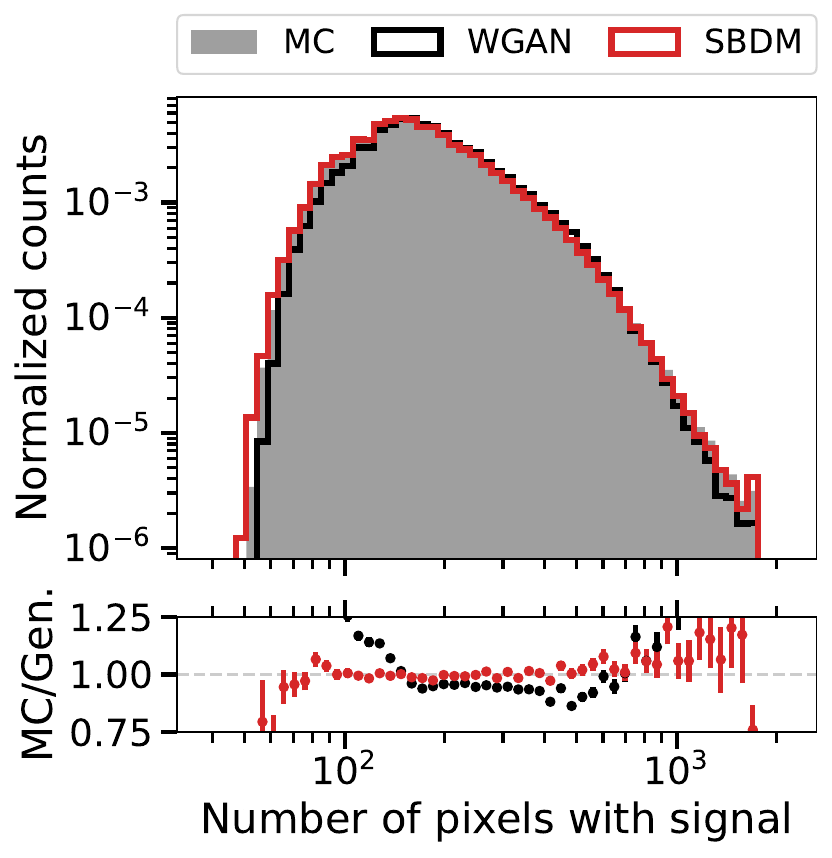}
  \includegraphics[scale=0.355]{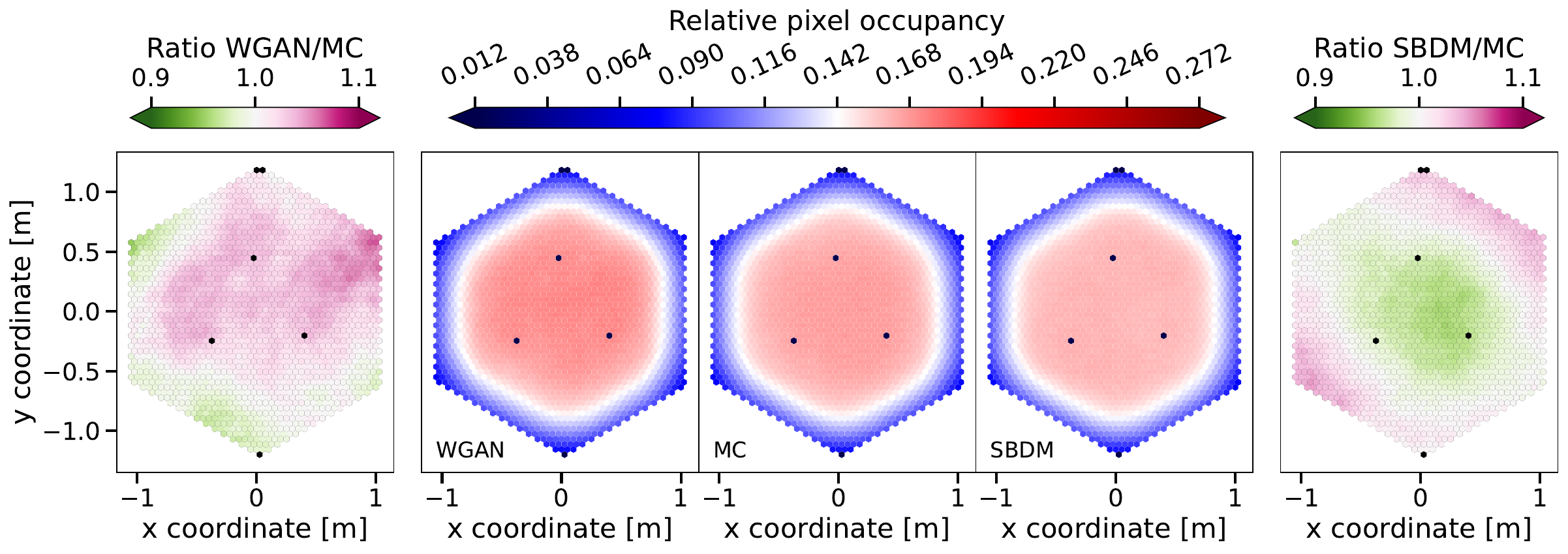}
  \caption{
  \label{fig:gamma_low_level}
  Comparison of the pixel values (top left), the image sizes (top middle), the number of pixels (top right), and the pixel occupancy (bottom) for the $\gamma$-ray dataset from the MC simulation, the WGAN, and the SBDM.
  The range of the color bar of the relative pixel occupancy is set to show $\pm 5\sigma$ around the mean value of the MC image.
  }
\end{figure}

\subsection{Low-level observables}
\label{sec:gamma_low_level}

Next, a quantitative analysis of the generated $\gamma$-ray images is performed to verify the generation of accurate IACT images.
In the following, four low-level observables are presented and discussed: each pixel value, the image size, the number of signal pixels in the image, and the relative pixel occupancy over the respective dataset.
The aim of this part is primarily to compare low-level image quantities, without yet looking into the more physical high-level parameters.
Figure~\ref{fig:gamma_low_level} compares the distributions for these low-level observables between the MC simulations (filled grey histogram), WGAN (black line) and SBDM (red line).

\paragraph{Pixel values and image size}
The histogram of individual pixel values is an important distribution, as it enables us to make a direct comparison between the generated p.e. distributions for the different approaches.
It ranges from about $-4$\,p.e. to 4176\,p.e. with the latter being constrained by the simulated saturation of the camera.
Both machine learning models are able to reproduce this distribution well, even though small deviations are observed at higher pixel values.
In particular, the WGAN shows deviations in the order of 5\,\%. 
This result demonstrates that the generative models can achieve near-complete reproduction of the phase space of the measured signals in the camera.

In the SBDM, the total image size is predicted by the global model and then passed to the image model as an additional input. The size values range from 140\,p.e. to almost $10^6$\,p.e. with the parameter distribution peaking at around 400\,p.e..
The image size is accurately represented in the images from the SBDM with small differences at the extreme size values. 
In the case of the WGAN, these differences are larger, with clear mismatches present below 600\,p.e. and for the highest values.

\paragraph{Number of signal pixels and pixel occupancy}

The number of signal pixels provides spatial information of the measured shower, in particular on the image structure close to the cleaning threshold (those pixels dominate the number).
This parameter ranges from 0 to 1758, the maximum number of operating PMTs in the camera.
The distribution of the number of pixels in the SBDM-generated images agrees well with that in the MC simulations, with only the first few bins showing small discrepancies.
Although the WGAN-generated images show poorer agreement in this quantity, their general trend follows that of the MC-generated images.

Lastly, the relative pixel occupancy quantifies the probability of a pixel containing a signal after cleaning, estimated over the whole dataset. This distribution is shown in figure~\ref{fig:gamma_low_level}, bottom. Most of the pixels in the inner part of the camera contain a signal in roughly 17\,\% of the images. The pixels close to the edge of the camera are occupied by less than 14\,\%, decreasing to a relative occupancy of less than 10\,\% at the edges.
The occupancy distributions of the WGAN and SBDM agree with the one from the MC-generated images, showing an accurate representation. For a more detailed comparison, we can look at the ratio of the occupancy images, which are shown on the left and right sides. Here, similarities and differences to the MC distributions are highlighted .
In the case of the SBDM, the occupancy in the middle of the camera is slightly higher than the average, and at the edges it is slightly lower, but remains within 5\,\%.
For the WGAN, the opposite is observed, with the occupancy being slightly higher in the middle and slightly lower at the edges.

In summary, the basic properties of the generated IACT images match those from the MC simulation, providing the first quantitative verification that the images are realistic.
Whereas the WGAN matches the quality from an earlier study~\cite{Elflein_2024}, the SBDM reproduces these parameters consistently with improved quality.

\subsection{High-level image parameters}
\label{sec:gamma_high_level}
Next, we continue with an analysis of high-level observables to quantify the physics properties of the generated images.
First, the air-shower impact point and the Hillas parameters, which are described in section~\ref{sec:benchmark}.
The parameters are obtained by using the roughly 98\,k IACT images in each dataset, which are left after we applied 9/16 tail-cuts cleaning and subsequent standard parameter cuts: minimum number of signal pixels of 10, minimum image size of 200\,p.e., and maximum radial coordinate of 0.8\,m.

\begin{figure}[t!]
  \centering
  \includegraphics[scale=0.45]{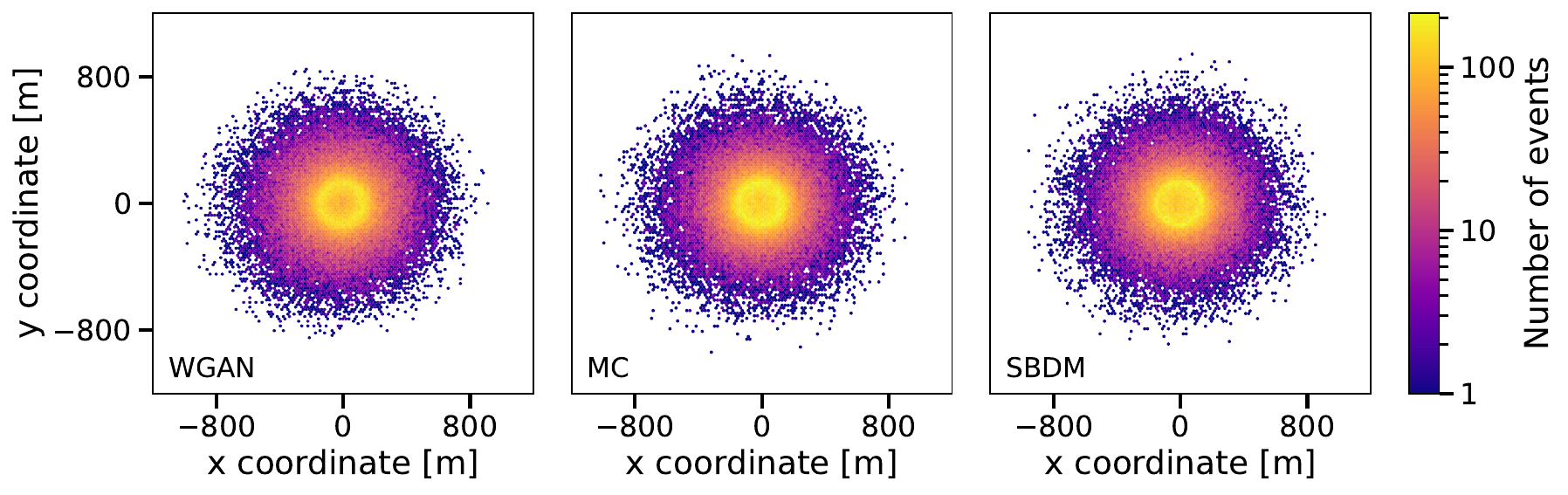}
  \caption{Visualization of the air shower impact points for the $\gamma$-ray data from the WGAN (left), the MC simulations (middle), and the SBDM (right). Each impact point corresponds to one image in the respective dataset. The Cherenkov telescope is located at the origin.}
  \label{fig:gamma_impact_point_map}
\end{figure}

\paragraph{Air-shower impact point}
The impact point is the location where the core axis of the  air shower intersects Earth's surface.
The \(x\) and \(y\) coordinates of the impact point are  predicted separately by the global model and given to the image model as additional information.
In figure~\ref{fig:gamma_impact_point_map}, the impact point maps from the MC simulations, and the WGAN and SBDM datasets are shown, where the CT5 telescope is placed at the origin $(0, 0)$.
As expected, most impact points are close to the telescope, and the density decreases farther from it.
At around 150\,m to 200\,m, we can see a ring of impact points with the highest density, which is comparable to the size of the Cherenkov cone.
We find nice agreements with the MC simulation for both DL models.

\begin{figure}[t!]
  \centering
  \includegraphics[scale=0.35]{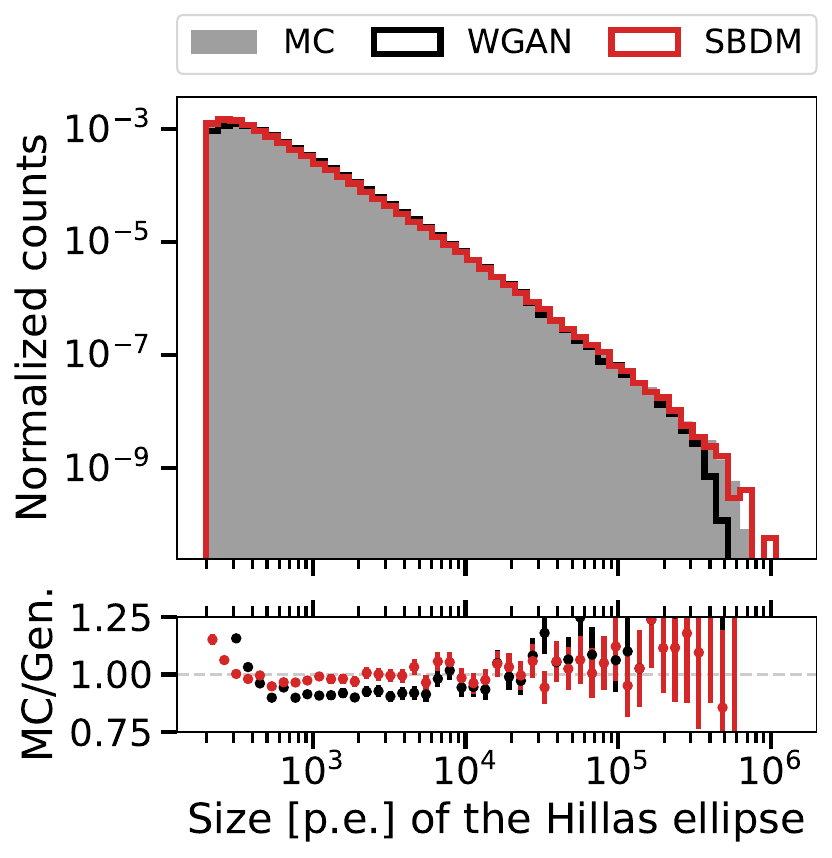}
  \includegraphics[scale=0.35]{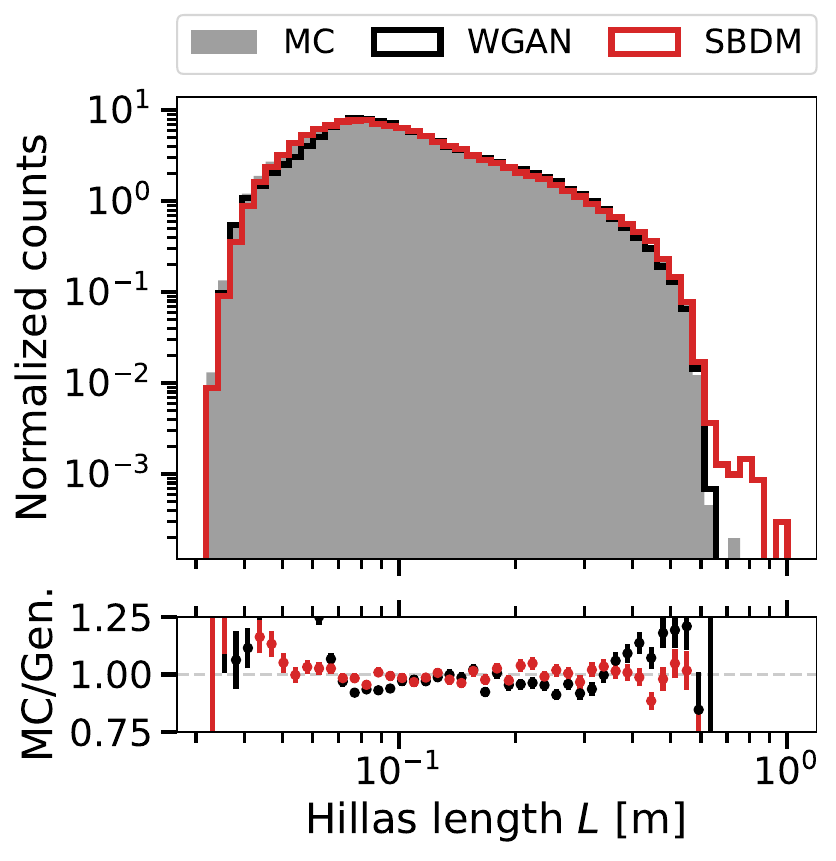}
  \includegraphics[scale=0.35]{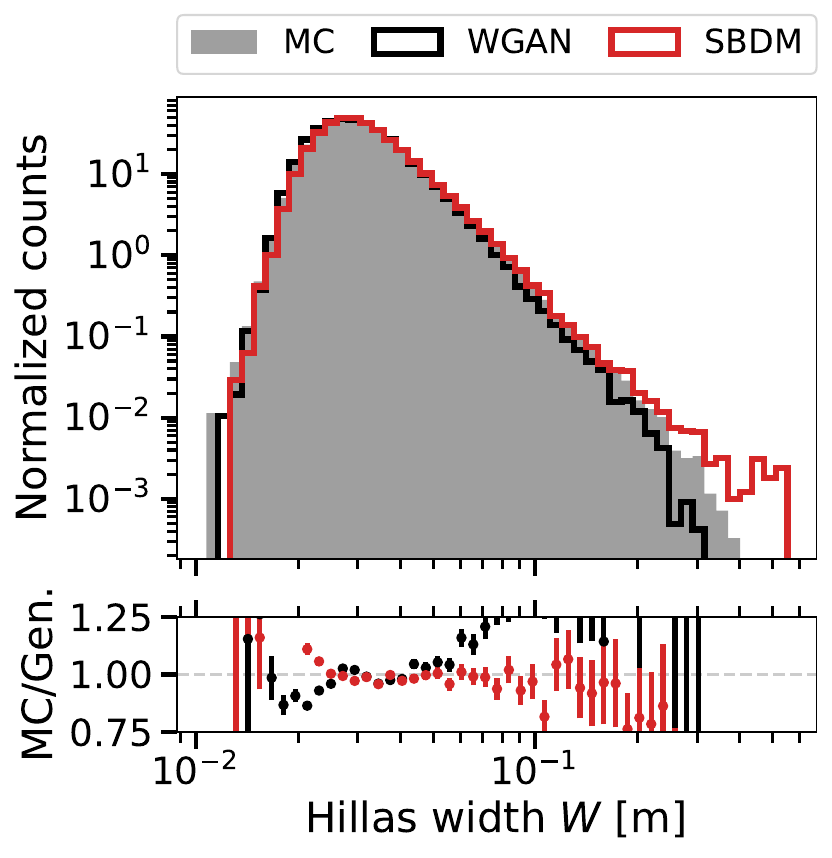}
  \includegraphics[scale=0.35]{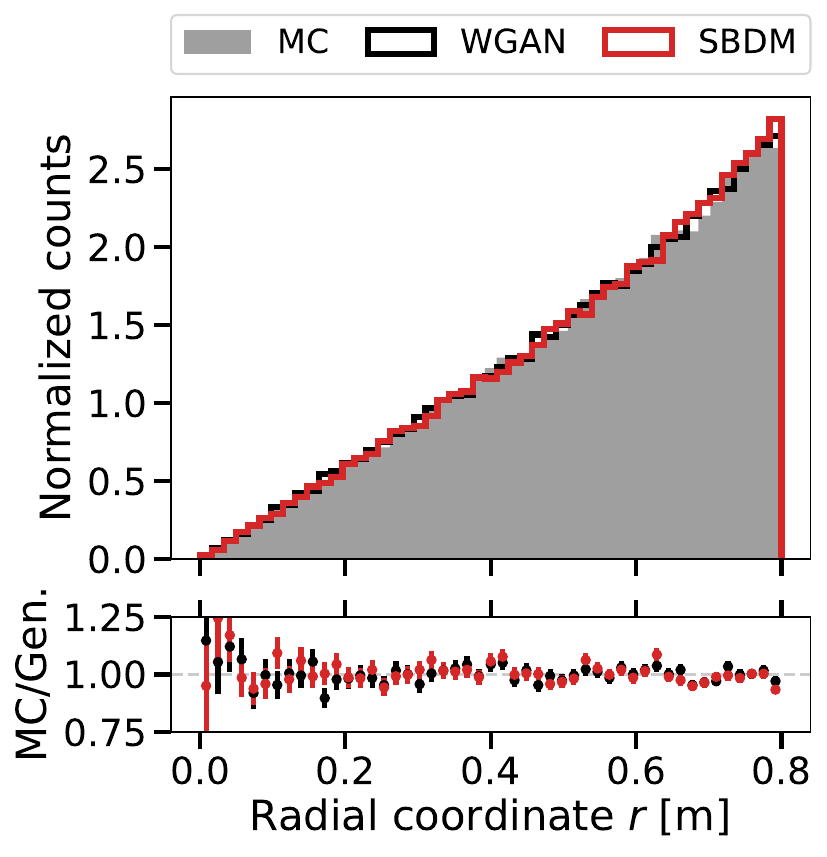}
  \includegraphics[scale=0.35]{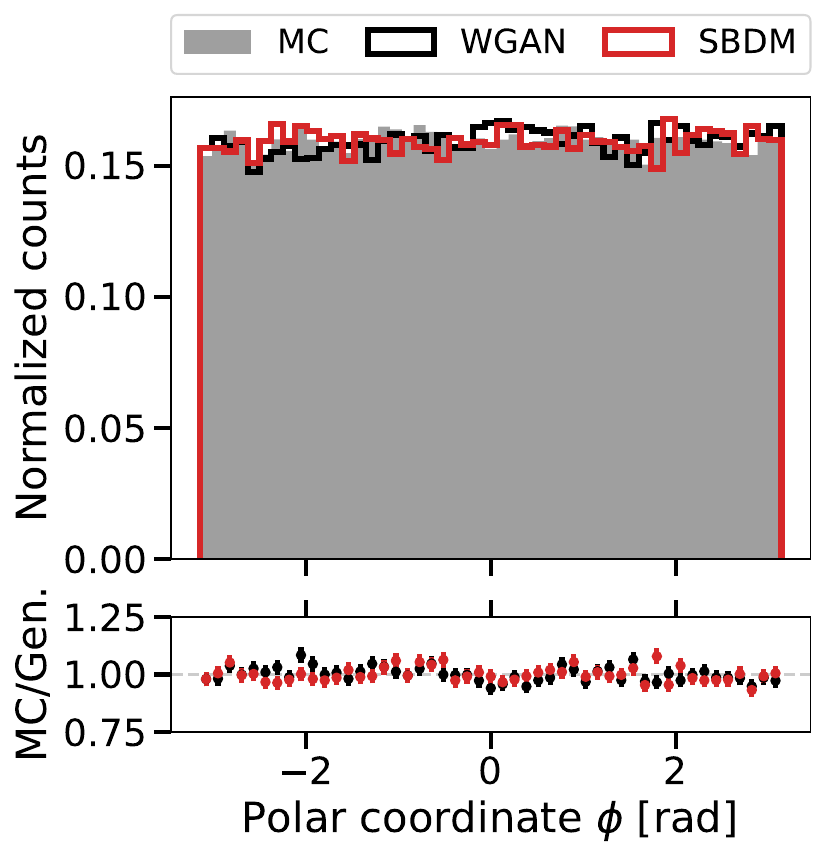}
  \includegraphics[scale=0.35]{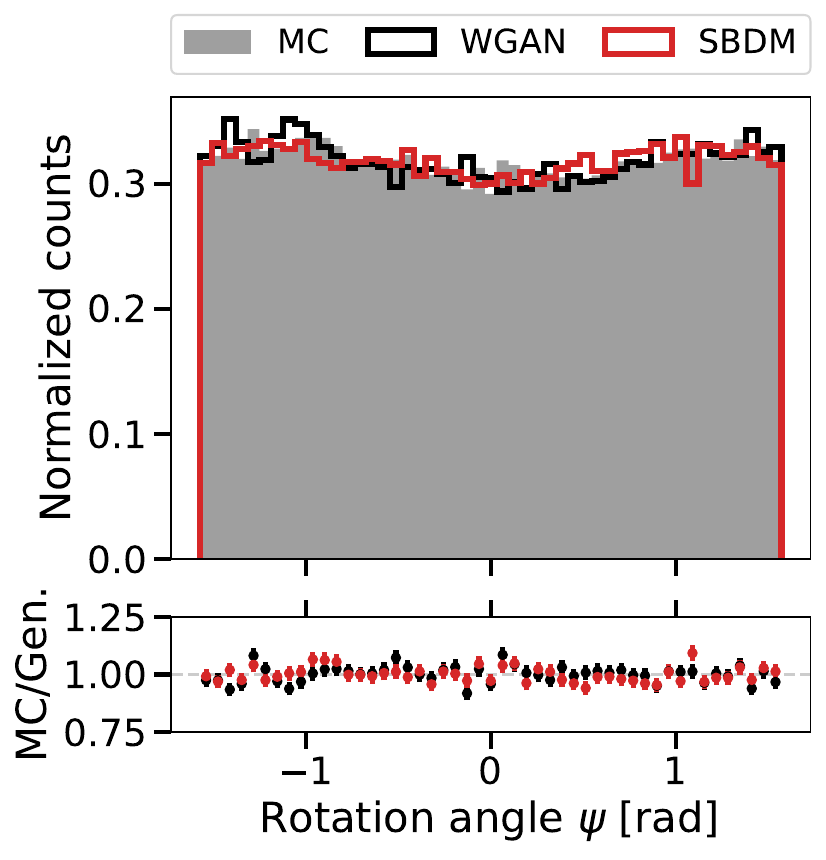}
  \includegraphics[scale=0.35]{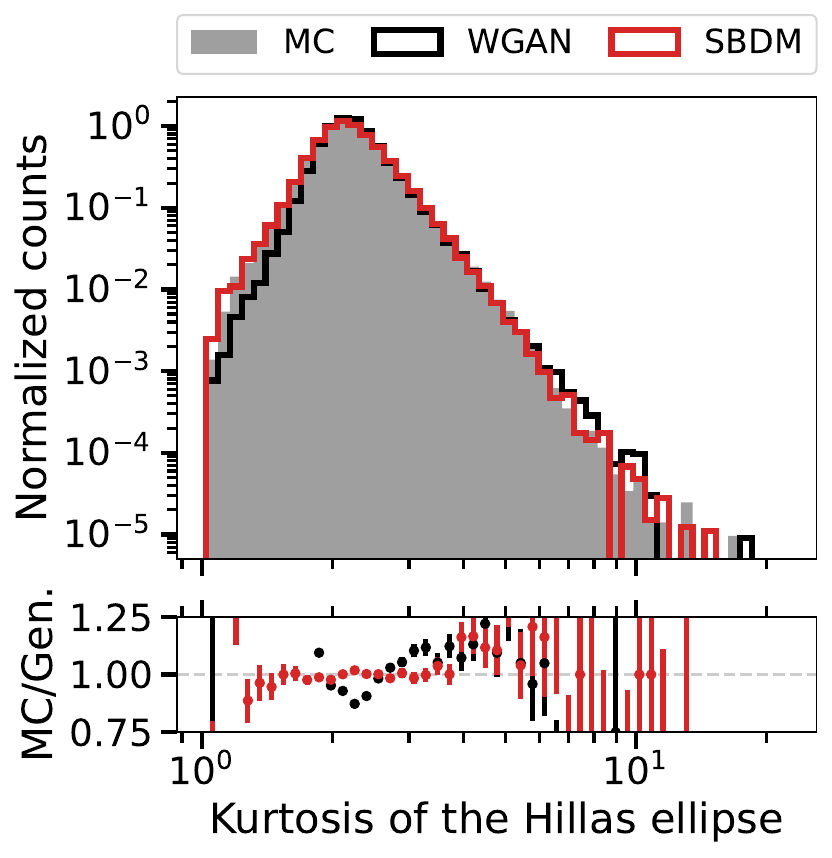}
  \includegraphics[scale=0.35]{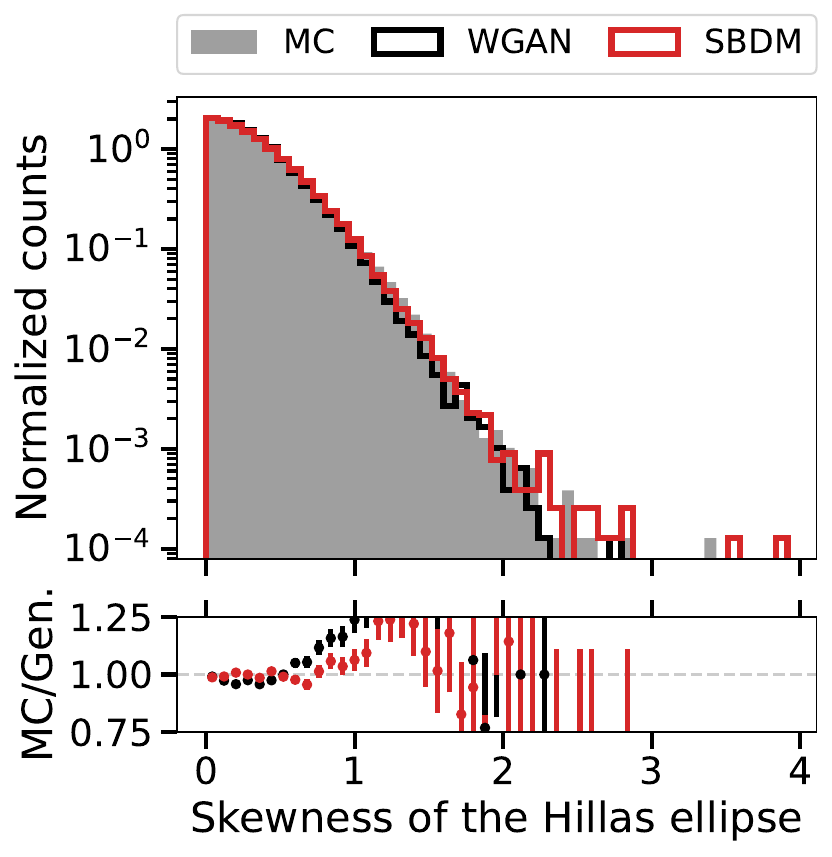}
  \caption{
  Histogram of Hillas parameters for the $\gamma$-ray dataset, comparing MC simulations (grey) with the generated WGAN (black), and generated SBDM (red) dataset.
  Below the distributions, the binned ratio of the MC data to the corresponding ML-generated dataset is displayed.
  Top: Hillas size, length, and width. Middle: Radial coordinate, polar coordinate, and rotation angle.
  Bottom: Kurtosis and absolute skewness of the Hillas ellipse.
  }
  \label{fig:gamma_hillas_params}
\end{figure}

\paragraph{Hillas size, length, and width}
The top row of figure~\ref{fig:gamma_hillas_params} shows the three key parameters: the Hillas size, the Hillas length, and the Hillas width.
Overall, a good agreement is observed across all distributions, indicating that most of the key parameters are modelled correctly in both approaches.
Looking at the distributions of Hillas size in detail, the WGAN generates fewer images at the highest Hillas size than the simulations, whereas the SBDM shows better agreement.
For the Hillas length, a great description is observed everywhere except at the highest lengths, where the SBDM generates more images than the simulations.
These artifacts are rather unlikely and cover a density of less than $4\cdot10^{-4}$ of the dataset if we consider lengths above 0.6\,m.
For the Hillas width, the SBDM similarly generates more events with the highest widths than simulations, whereas the WGAN generates fewer events with the highest widths.
Although the SBDM shows slight improvements over the WGAN within the bulk of the distribution, more research is foreseen for investigating the limitations for large length and width images in the future.

\paragraph{Radial coordinate, polar coordinate, and rotation angle}
The next three Hillas parameters to examine, which provide information on the position of the signal in the camera, are shown in the second row of figure~\ref{fig:gamma_hillas_params}.
All of the  geometric parameters studied for the WGAN and the SBDM, comprising the radial coordinates (left), polar coordinate ranges (center), and the rotation angle (right), exhibit excellent agreement with the simulation.

\paragraph{Kurtosis and skewness}
Finally, the kurtosis and the absolute value of the skewness of the Hillas ellipse are studied in the bottom row of figure~\ref{fig:gamma_hillas_params}.
These higher-order parameters are the most challenging to learn.
Nonetheless, aside from small deviations at the extremes of the distributions, the generated datasets match the MC well for the bulk of the distribution, with the SBDM featuring small improvements.\newline

In summary, all studied high-level parameters are reproduced well by the deep-learning-based approaches, with the SBDM showing slightly better agreement across most of the significant parts of the phase space.
At extreme regions, small discrepancies are observed, possibly due to a lack of examples with extreme values.

\subsection{Correlations}
\label{sec:gamma_correlations}

\begin{figure}[t!]
  \centering
  \includegraphics[scale=0.4]{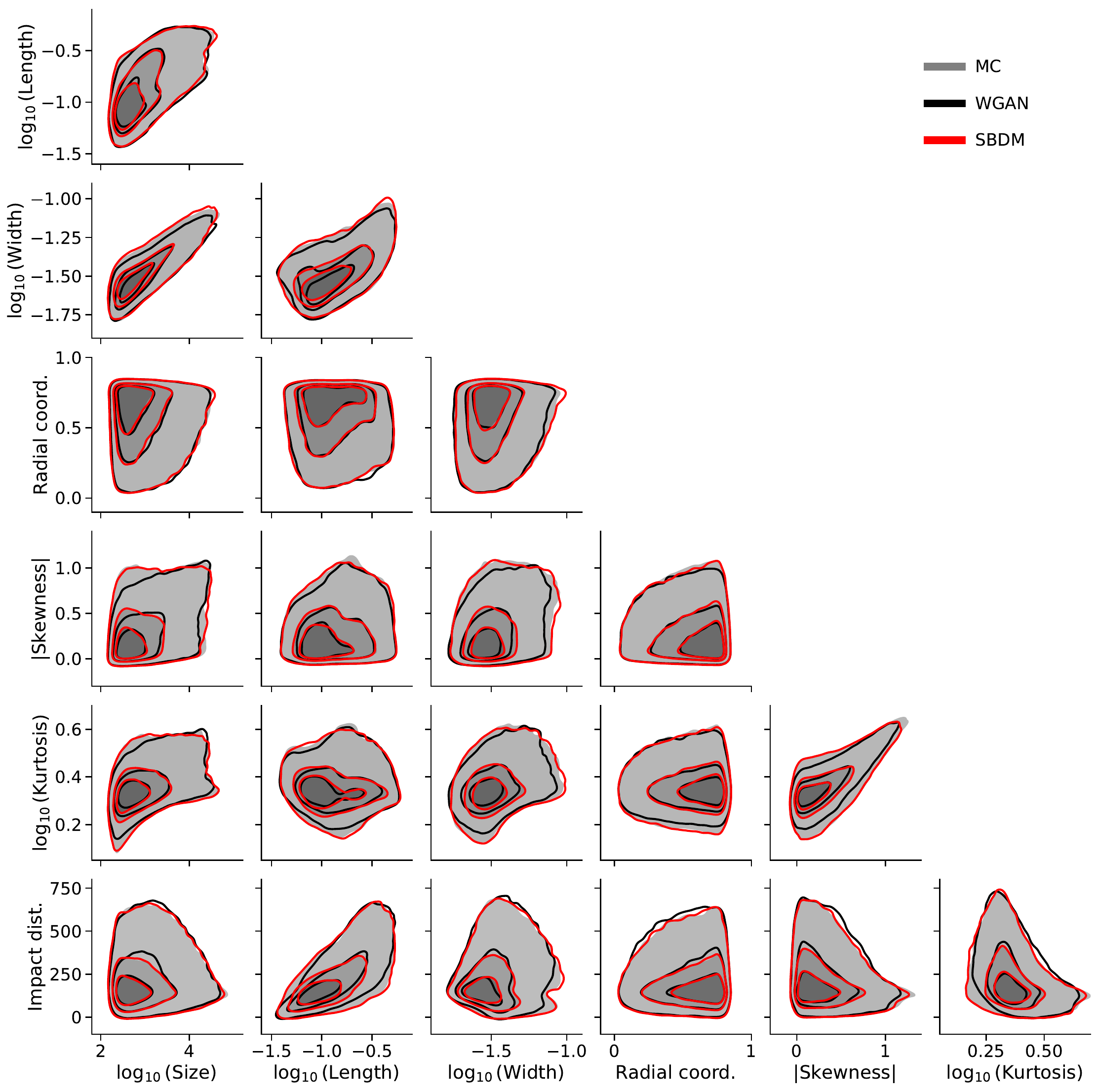}
  \caption{
  Correlations between the air shower impact points and Hillas parameters of the MC (grey), WGAN (black), and SBDM (red) $\gamma$-ray dataset.
  The correlations are displayed as kernel density estimation levels with the inner regions corresponding to higher densities.
  } \label{fig:gamma_correlations}
\end{figure}

In the final part of this study, we examine the correlations between high-level parameters.
For this investigation, we highlight the correlations between the most complex and relevant distributions, comprising size, length, width, radial coordinate, skewness, kurtosis, and impact distance\footnote{The impact distance is defined as the distance between the impact point and the telescope position on the ground.}.
The correlations of these parameters, created visually using the four-level Kernel Density Estimation (KDE), for the MC, WGAN, and SBDM $\gamma$-ray dataset are shown in figure~\ref{fig:gamma_correlations}.
For the MC data, the correlations are shown as filled regions, with intensity reflecting the density.
In the case of the WGAN and the SBDM the correlations are visualized through black and red lines, which are the boundaries of the different KDE levels.
While the SBDM is overall in remarkably good agreement with the MC, some minor deviations are observed in low-density regions, for example, between skewness and width. In contrast, significant deviations are observed between the MC and WGAN, especially, kurtosis and skewness exhibit more noticeable deviations.

While the visual representation of the correlations already gives a very good estimate of the accuracy of the models, we performed a quantitative analysis for a proper comparison with the used metrics being explained in detail in section \ref{sec:benchmark}.
We first estimate the Frobenius distance using the impact distance and the six chosen Hillas parameters shown in figure~\ref{fig:gamma_correlations}, as well as compute the correlation matrix similarity.
Furthermore, we perform a machine-learning-based classification between the MC, WGAN, and SBDM datasets using a random forest.
The results of each test are summarized in table~\ref{tab:correlation_gamma}.
In addition, we show the optimal values achievable.
In all cases, we find significantly improved performance of the SBDM compared to the WGAN.

Regardless, both models are well-suited for generating high-quality $\gamma$-ray images, and their performance shows only small discrepancies.
The WGAN is capable of generating high-quality IACT $\gamma$-ray images, as demonstrated in this work and in ref.~\cite{Elflein_2024}, reproducing both low-level and high-level parameter distributions with high fidelity and preserving their correlations with acceptable quality.
The SBDM developed in this work achieves very good precision for most observables; however, overall the performance differences are modest and most striking for the correlations, where the SBDM is superior.

\begin{table}[t!]
\centering
\begin{centering}
\begin{tabular}{ c | c | c | c  }
\hline
& {Frobenius} & {Correlation matrix} & {Accuracy} \\
& {distance} & {similarity} & {random forest} \\
\hline
Optimal case & 0 & 1 & 0.5 \\
MC vs. WGAN & $0.36 \pm 0.01$ & $0.9914 \pm 0.0004$ & $0.678 \pm 0.004$ \\
MC vs. SBDM & $0.09 \pm 0.01$ & $0.9991 \pm 0.0002$ & $0.515 \pm 0.005$ \\
\hline
\end{tabular}
\caption{
Quantitative performance measures: Frobenius distance, correlation matrix similarity, and random forest accuracy for the generated $\gamma$-ray datasets comparing the WGAN and the SBDM with its similarity to MC. As a reference, the optimal values using indistinguishable datasets are shown in the first row.}
\label{tab:correlation_gamma}
\end{centering}
\end{table}

\begin{figure}[t!]
  \centering
  \includegraphics[scale=0.37]{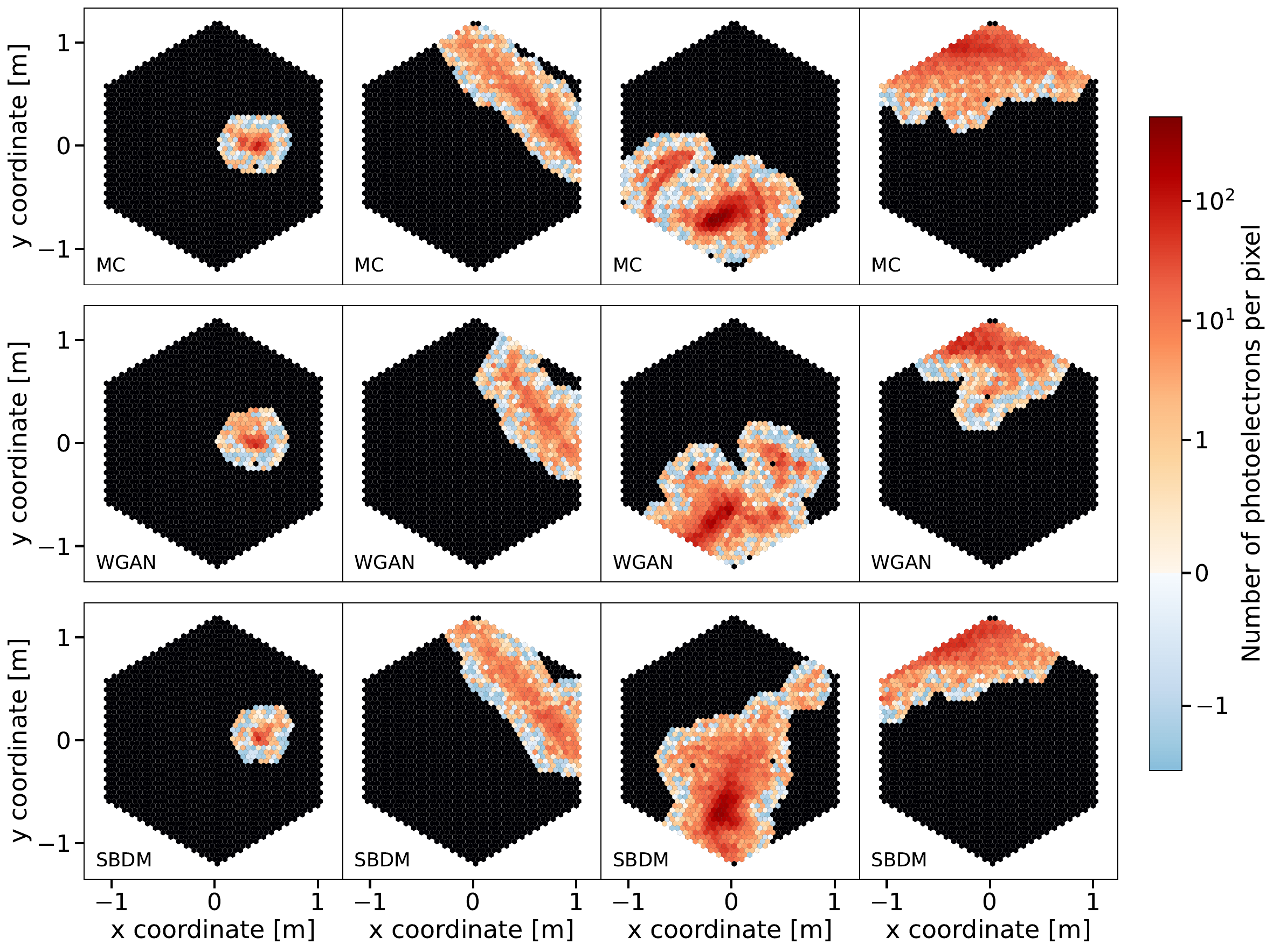}
  \caption{Comparison of four proton IACT images obtained from MC simulations (top), the WGAN (middle), and the SBDM (bottom).
  The simulated images are hand-picked to show various characteristics of air showers induced by protons.
  The WGAN and SBDM images are the next neighbors of the simulated images in the mean-squared-error pixel space.
  Note that a 4/7 tail-cuts cleaning with the extension of four rows is applied, and black pixels denote pixels that do not contain any signal.}
  \label{fig:proton_images}
\end{figure}

\section{Analysis of generated proton images}
\label{sec:proton}
In the following section, the generation quality of proton showers using the WGAN and the SBDM is investigated.
Due to the inherently complex nature of hadronic air showers and thus proton events, the following study provides insight into the modelling of more complex IACT images.
The MC, WGAN and SBDM datasets after cleaning contain approximately 93\,k, 76\,k, and 93\,k post-processed images, respectively.
As before, the analysis starts with a short visual investigation of hand-picked simulated images in section~\ref{sec:proton_visual}, followed by studying low-level observables in section~\ref{sec:proton_low_level}.
Lastly, the Hillas parameters, as well as their correlations, are examined in section~\ref {sec:proton_high_level} and section~\ref{sec:proton_correlations} to verify the accurate reproduction of the inherent physics properties of the generated IACT proton images.

\subsection{Visual inspection}
\label{sec:proton_visual}
The visual quality of the generated proton images is shown in figure~\ref{fig:proton_images}, depicting four distinct events.
The simulated images are handpicked to show the various air-shower characteristics, and the next neighbors in MSE-pixel space for the SBDM and the WGAN are shown as a comparison.
In the first two columns, images with circular and elliptical signals can be seen.
While such image types are partly produced in the hadronic component of air showers, they can also originate from electromagnetic sub-showers initiated by the decay of neutral pions.
The morphology in the images shown in the last two columns is considerably more irregular, i.e., contains more hadronic substructure, in particular compared to the $\gamma$-ray images (cf. figure~\ref{fig:four_gamma}).
Due to this irregularity, identifying generated images that closely match the selected examples is challenging.
Nevertheless, the generated images reproduce the characteristic features expected from cosmic-ray–induced air showers, indicating that both the WGAN and the SBDM are capable of generating IACT images that appear realistic to the human eye.

\begin{figure}[t!]
  \centering
  \includegraphics[scale=0.25]{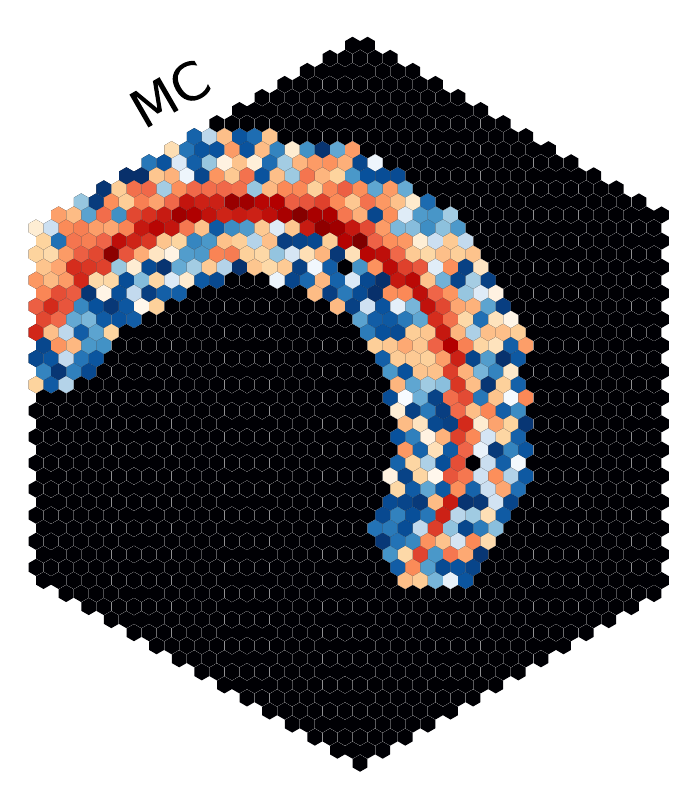}
  \includegraphics[scale=0.25]{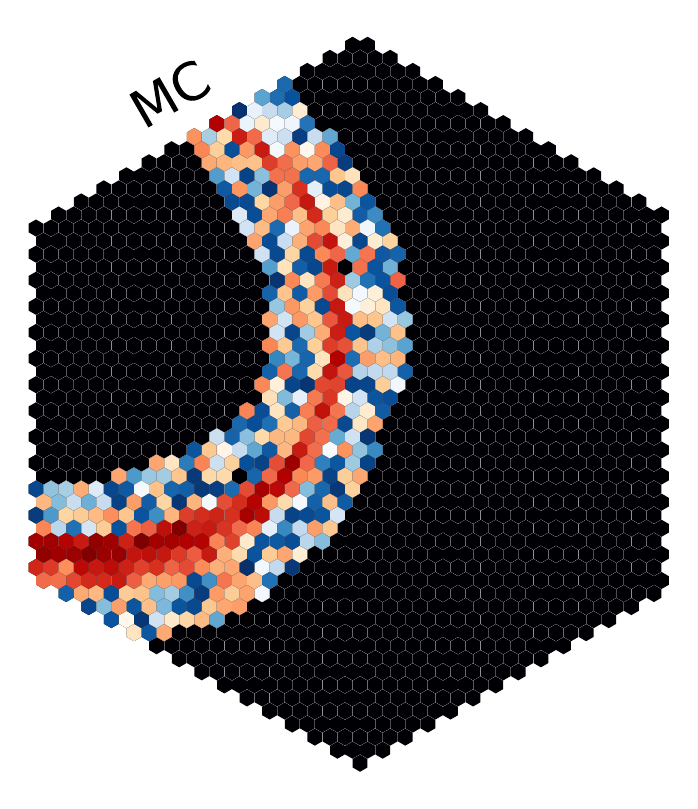}
  \includegraphics[scale=0.25]{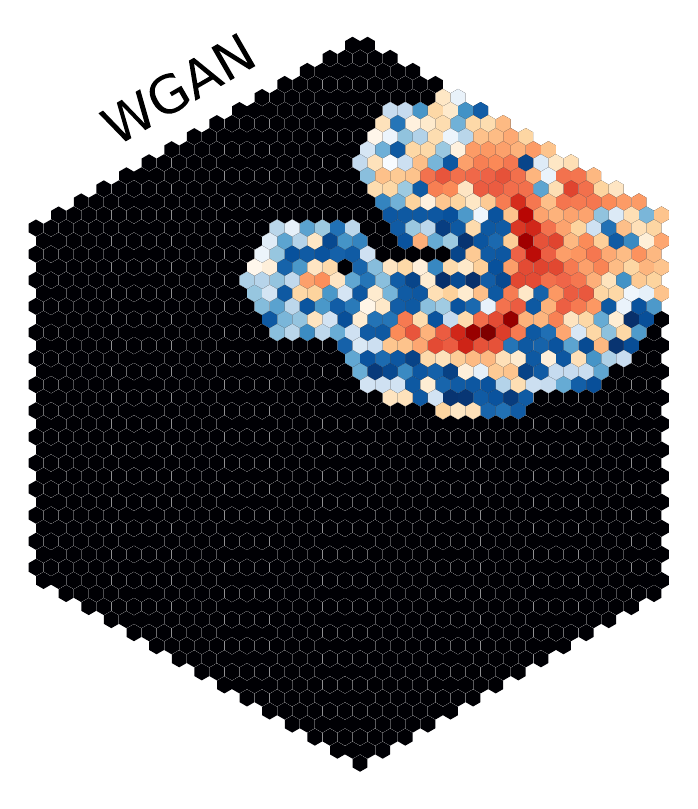}
  \includegraphics[scale=0.25]{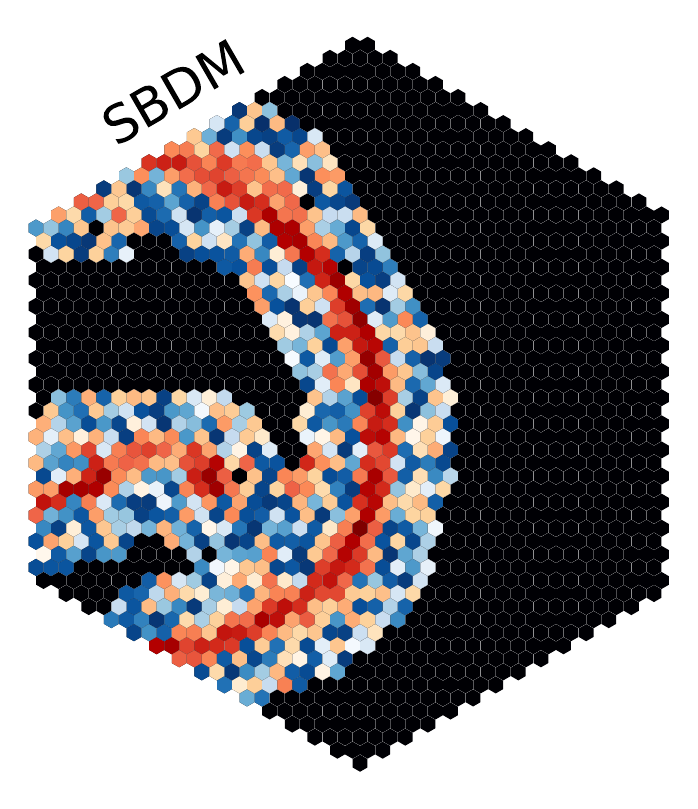}
  \includegraphics[scale=0.25]{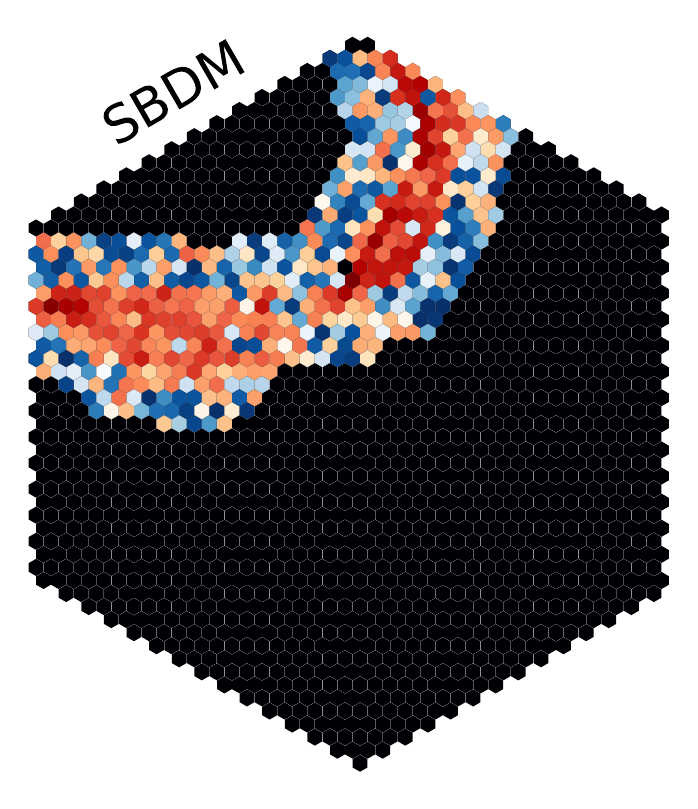}
  \caption{IACT images containing muon rings.
  The first two images are from the simulated proton dataset; the third is from the WGAN data; and the other two are from the SBDM data.
  The dark-red pixels correspond to large-signal pixels, which contain most of the detected Cherenkov light.
  Black pixels indicated cleaned signals.
  Only the SBDM data contains distinct rings.}
  \label{fig:muon_rings}
\end{figure}

\paragraph{Complex images}
While almost all samples in the proton dataset are made of air-shower images, single muon events induce a distinctive circular pattern in the camera, so-called muon rings, when directly hitting the telescope.
Whereas the majority of the muon images contain only a part of a ring, as they do not cross the optical plane of the IACT orthogonally, only very few `full muon rings' show clear rings, which are, however, often superpositioned by a shower.
This very low density in the training data makes it hard to be learned by a generative model.
Excitingly, we find characteristic events with complex ring shapes in the proton data of the SBDM.
A collection of such images is shown in figure~\ref{fig:muon_rings}, with the first two images belonging to the MC proton data and the last two images being generated by the diffusion model.
While we were able to spot some realistic muon rings in the SBDM proton data, we did not find any clear muon rings in the WGAN proton data.
The only image shown shows a partly curved signal, which shares some similarities with a muon ring, making the SBDM qualitatively superior to the WGAN in terms of visual image characteristics.
As tagging muon rings in IACT images~\cite{tyler2013muonidentificationveritasusing, Feng_2016, chalmecalvet2014muon} is challenging and research on their reliable identification is ongoing, investigating their overall density is beyond the scope of this work.

\subsection{Low-level observables}
\label{sec:proton_low_level}

\begin{figure}[t!]
  \centering
  \includegraphics[scale=0.35]{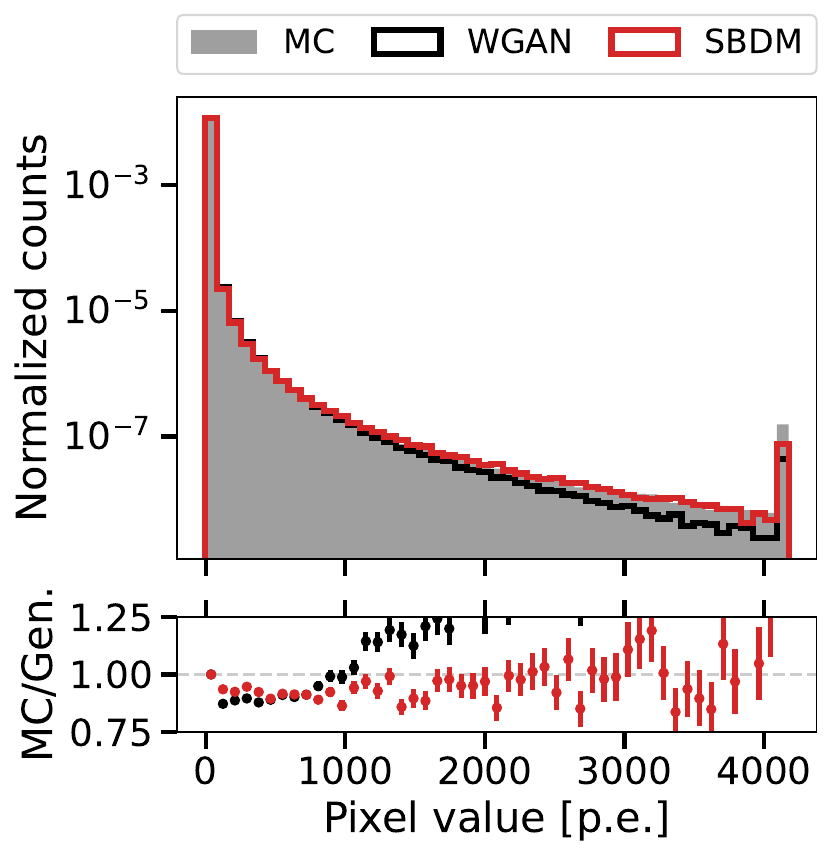}
  \includegraphics[scale=0.35]{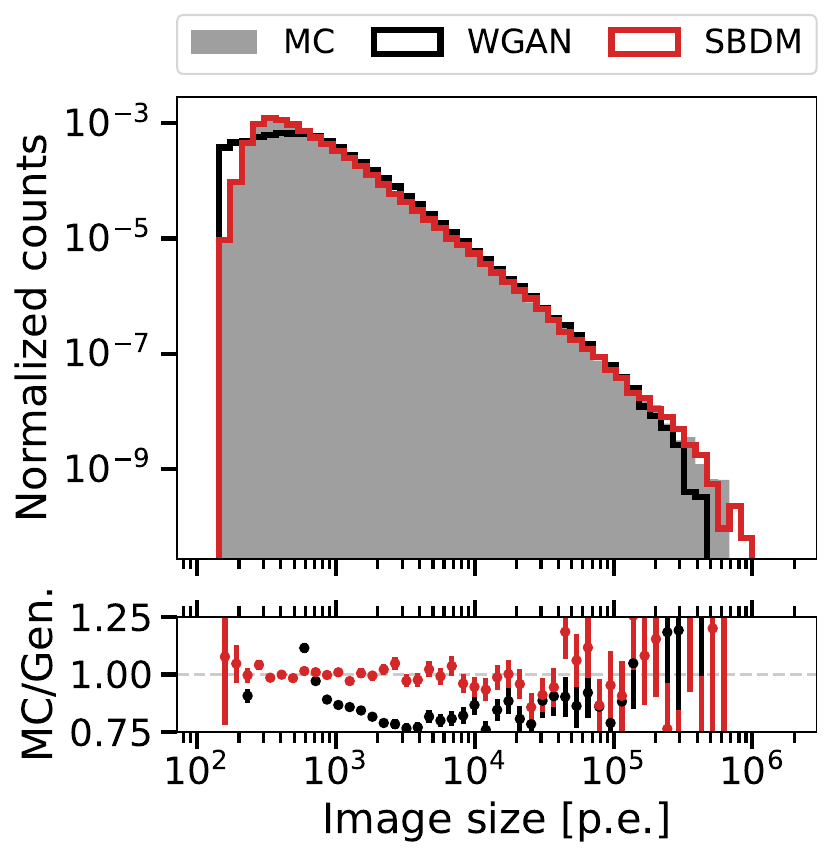}
  \includegraphics[scale=0.35]{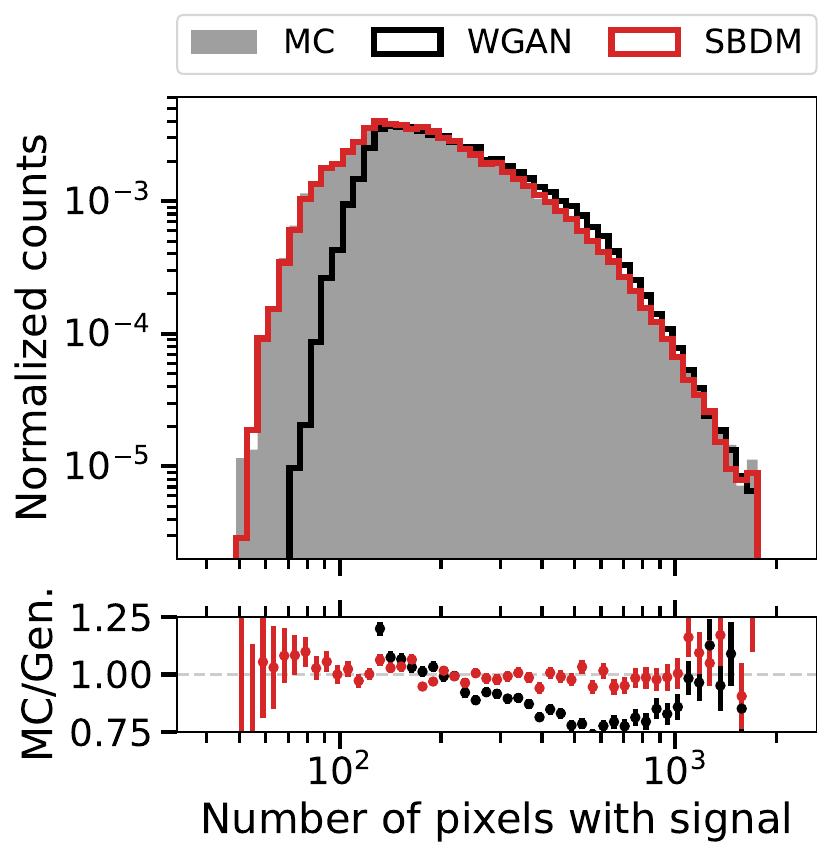}
  \includegraphics[scale=0.355]{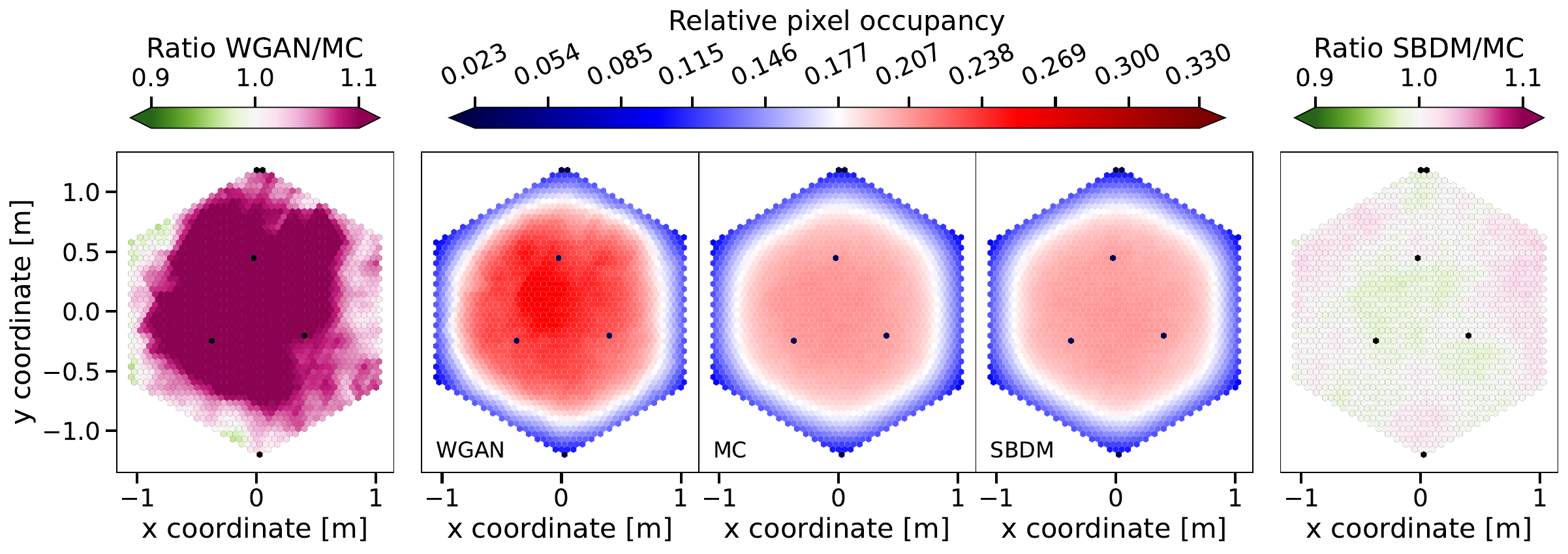}
  \caption{
  Comparison of the pixel values (top left), the image sizes (top middle), the number of pixels (top right), and the pixel occupancy (bottom) for the proton dataset from the MC simulation, the WGAN and the SBDM.
  The range of the color bar of the relative pixel occupancy is set to show $\pm 5\sigma$ around the mean value of the MC image.
  }
  \label{fig:proton_low_level_plots}
\end{figure}

After a qualitative inspection of the images, a quantitative analysis of the low and high-level parameters is performed, following a similar approach as done for the $\gamma$-ray images.
The low-level parameters, comprising pixel values (left), image size (center), and number of pixels after cleaning (right) are shown in figure~\ref{fig:proton_low_level_plots} together with the occupancy (bottom), i.e., the probability of a pixel containing signal after cleaning.

\paragraph{Pixel values and image size}
Similar to the $\gamma$-ray case, the full phase space from $-3$\,p.e. to about 4176\,p.e. (limited by saturation) is covered by the two generative models.
The distributions of the simulated and SBDM data match well, with some differences only present at smaller sizes.
In contrast, the WGAN shows significant deviations in the order of 20\,\% and more for pixel values about 1000 p.e..
A similar behavior can be seen for the image size, which ranges from roughly 140\,p.e. to $10^6$\,p.e..
Whereas the WGAN can not precisely reproduce the simulated distribution, the SBDM shows excellent agreement.
This is a significant difference compared to the $\gamma$-ray case, indicating distinct limitations of the WGAN.

\paragraph{Number of signal pixels and pixel occupancy}
The limitations of the WGAN are even more significant for the number of pixels and the occupancy (cf. figure~\ref{fig:proton_low_level_plots} right and bottom).
While the SBDM data match the simulations well, clear differences are evident in the case of the WGAN, particularly close to the edges of the phase space in the image size, as well as a significantly increased occupancy in the center of the camera.
While the WGAN fails to reproduce the proton showers with high quality, the SBDM provides high quality.
These are interesting findings that underscore the complexity of modeling the evaluation of hadronic showers and highlight the limits of the WGAN approach.

\begin{figure}[t!]
  \centering
  \includegraphics[scale=0.475]{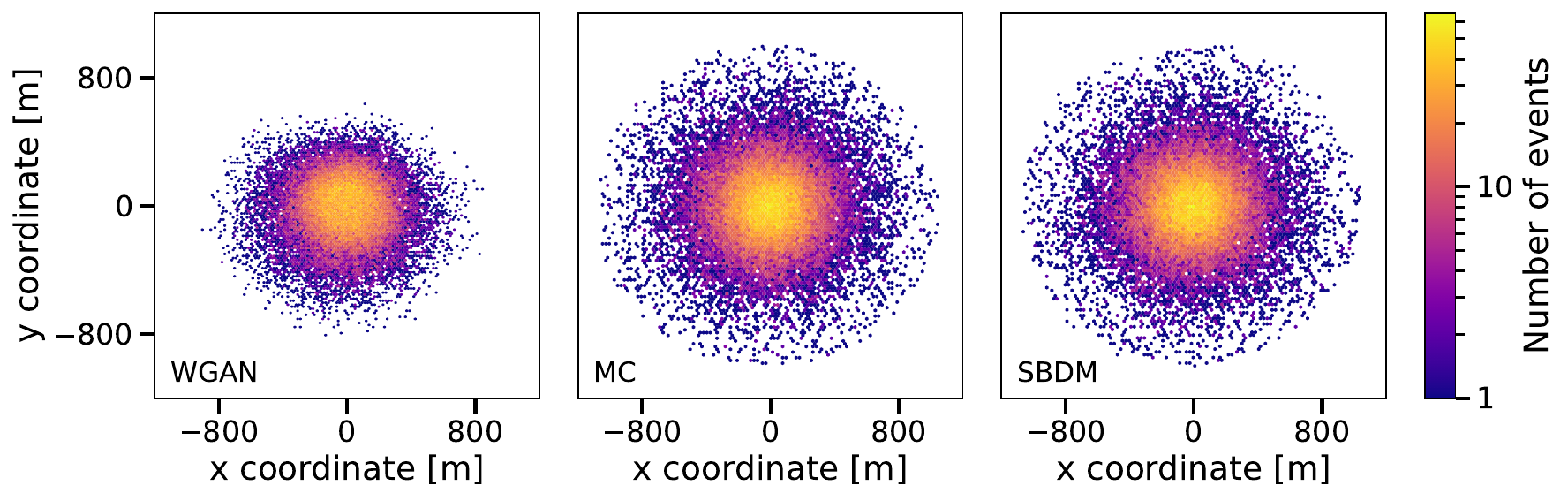}
  \caption{Visualization of the air shower impact points for the proton data from the WGAN (left), the MC simulations (middle) and the SBDM (right). Each impact point corresponds to one image in the respective dataset. The Cherenkov telescope and its camera is located at (0, 0).}
  \label{fig:proton_impact_point_map}
\end{figure}

\subsection{High-level image parameters}
\label{sec:proton_high_level}

As the next part of the image analysis, a comparison of various high-level parameters is carried out. First, the air-shower impact point predicted by the global model and used as input for the image model is investigated. In addition, the Hillas parameters, which are described in detail in section~\ref{sec:benchmark}, are examined.
As in the $\gamma$-ray analysis, we apply 9/16 tail-cuts cleaning and perform standard H.E.S.S.\ analysis cuts: minimum number of signal pixels of 10, minimum size of 200\,p.e., and maximum radial coordinate of 0.8\,m.
This leaves us with roughly 35\,k images for each dataset.

\paragraph{Air shower impact point} 
The map of air-shower impacts is shown in figure~\ref{fig:proton_impact_point_map}, with the CT5 telescope located at $(0, 0)$.
For both the MC and SBDM, the impact points reach from about $-1000$\,m to about $1000$\,m in both directions, and they are generally in good agreement.
However, this is not the case for the WGAN map, as the distribution is truncated and limited to about 800\,m.
Interestingly, this was not the case for the $\gamma$-ray dataset (see figure~\ref{fig:gamma_impact_point_map}).

\begin{figure}[t!]
  \centering
  \includegraphics[scale=0.345]{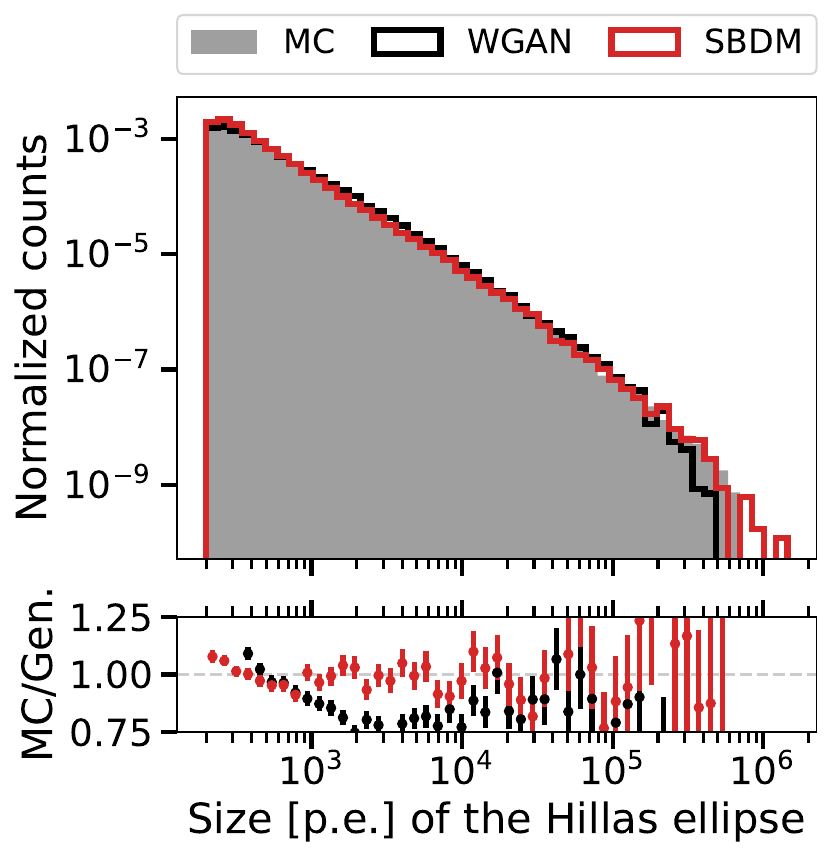}
  \includegraphics[scale=0.345]{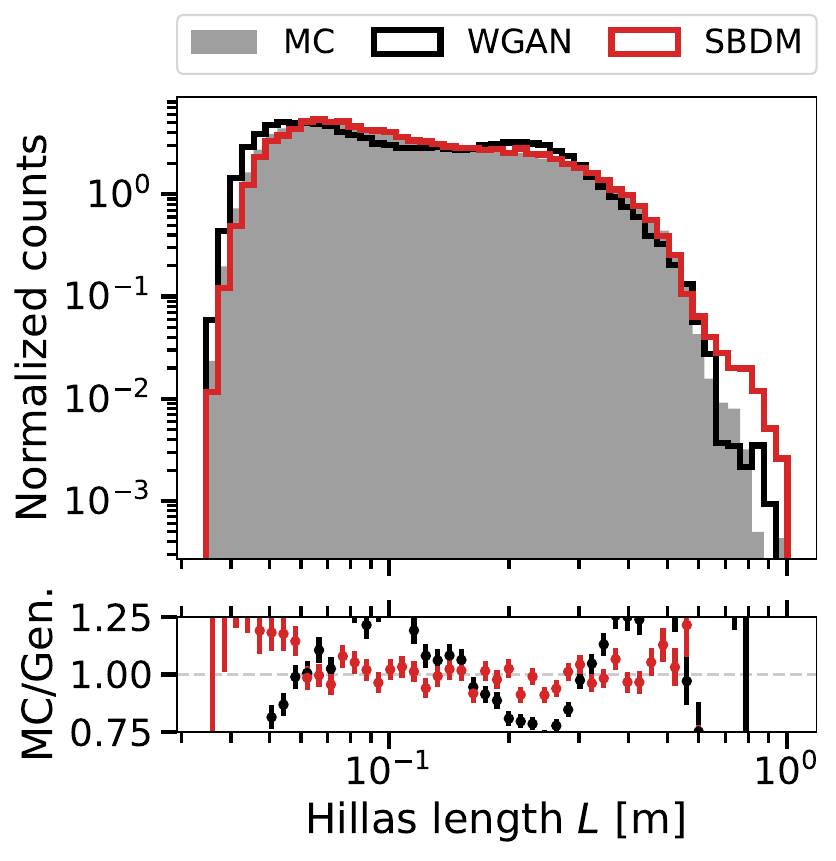}
  \includegraphics[scale=0.345]{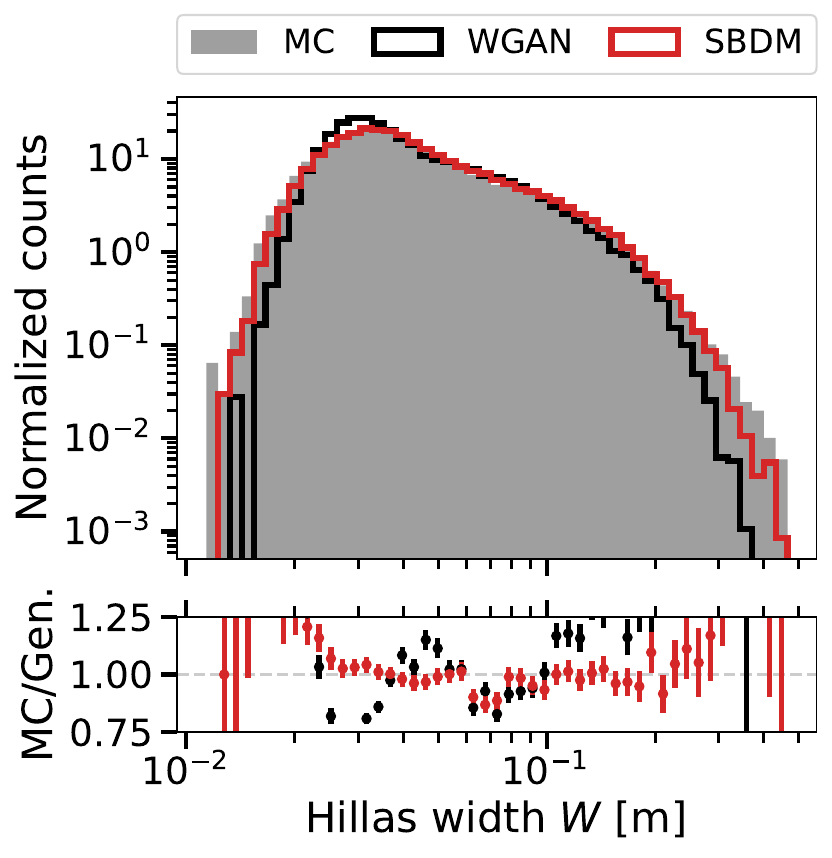}\\
  \includegraphics[scale=0.345]{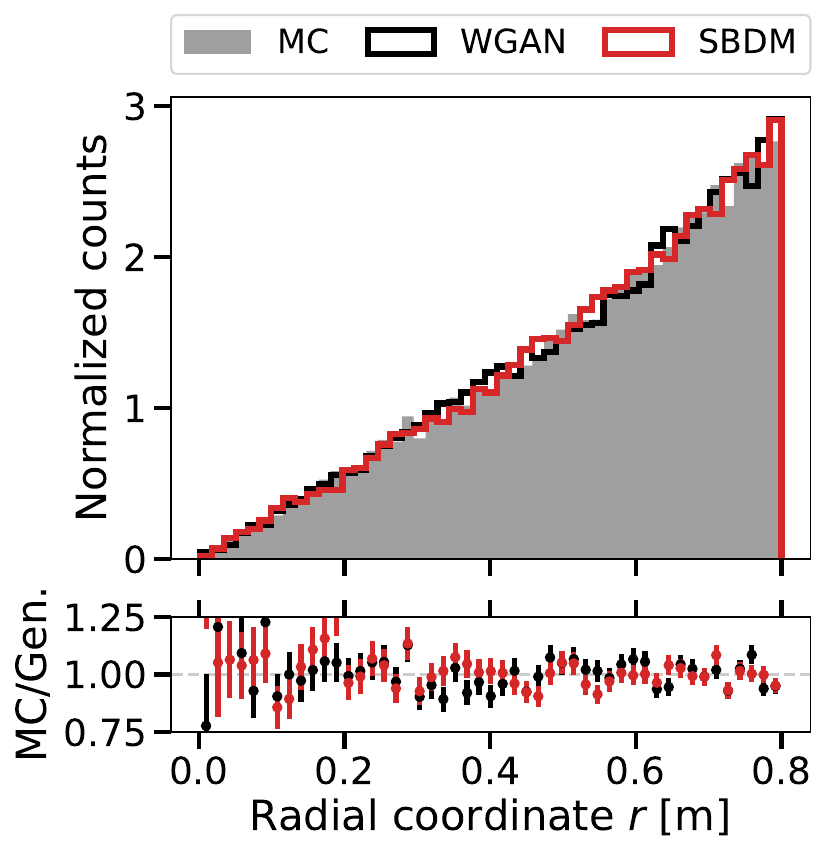}
  \includegraphics[scale=0.345]{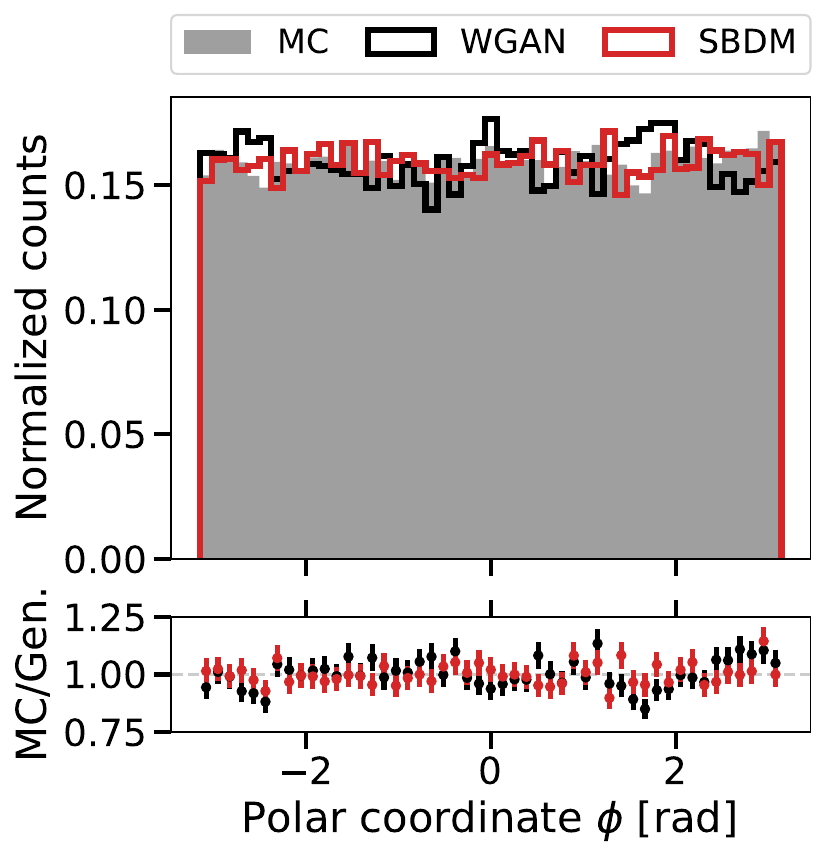}
  \includegraphics[scale=0.345]{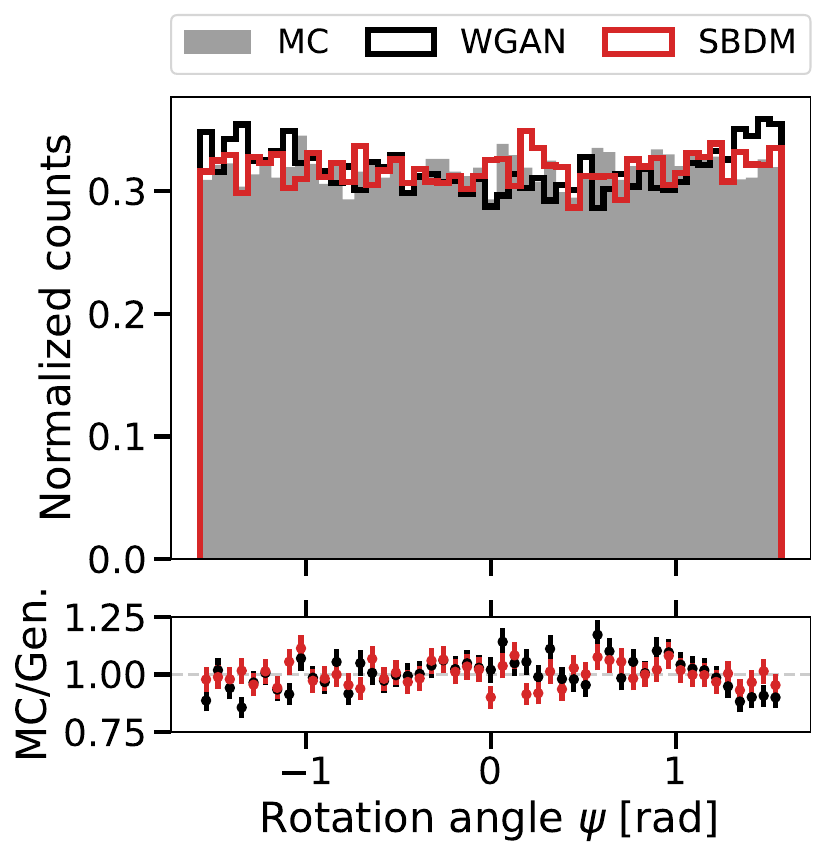}\\
  \includegraphics[scale=0.345]{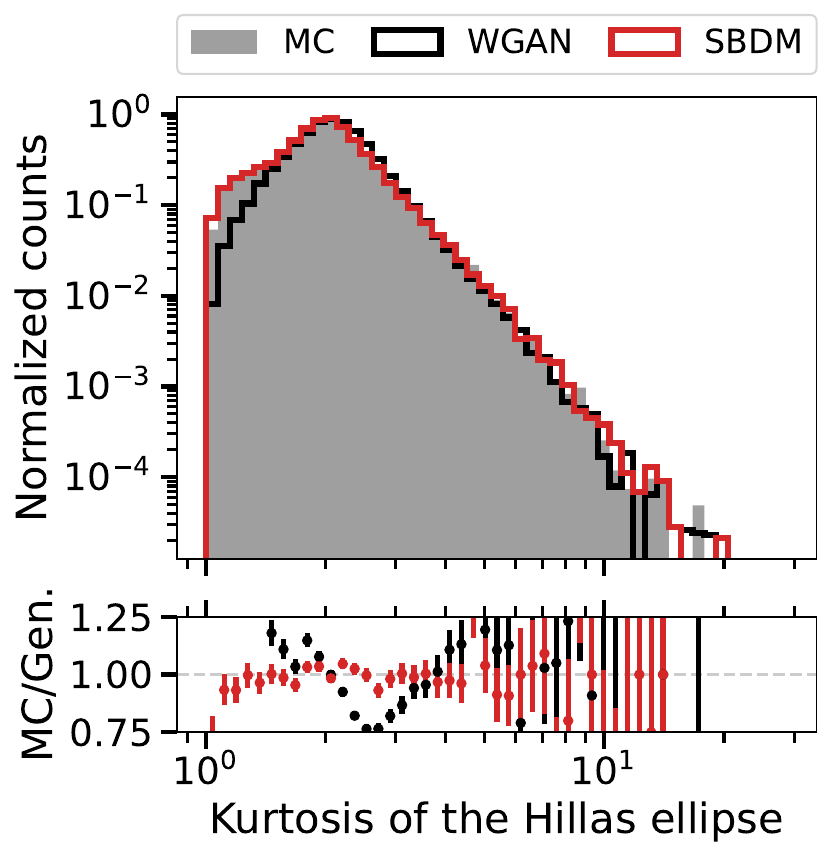}
  \includegraphics[scale=0.345]{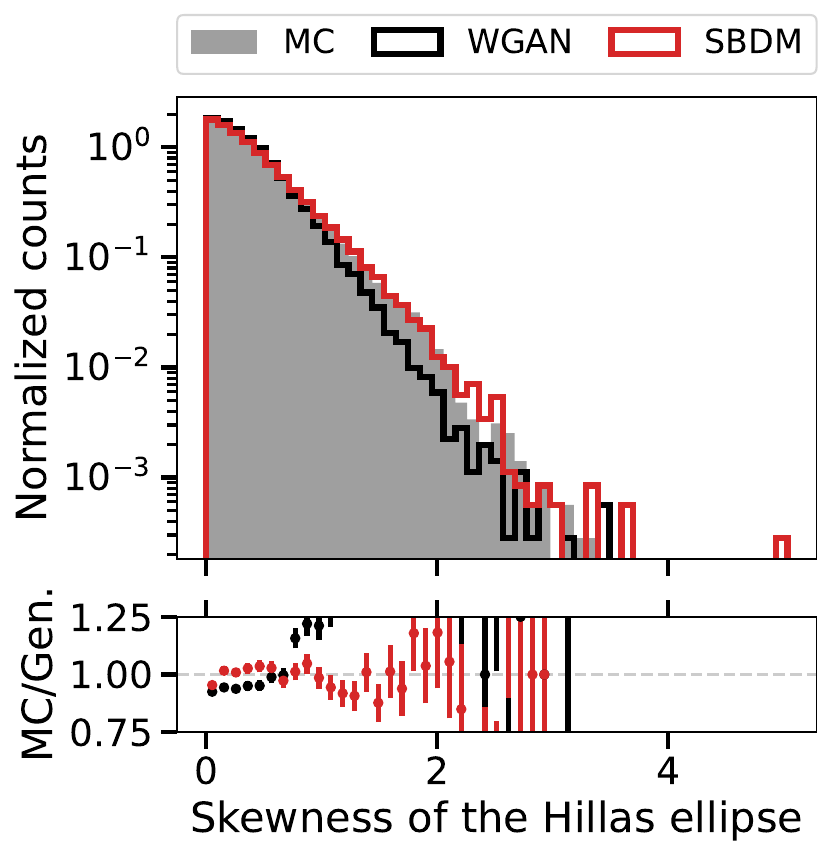}
  \caption{
  Histogram of Hillas parameters for the proton dataset, comparing MC simulations (grey) with the generated WGAN (black), and generated SBDM (red) dataset.
  Below the distributions, the binned ratio of the MC data to the corresponding ML-generated dataset is displayed.
  Top: Hillas size, length, and width. Middle: Radial coordinate, polar coordinate, and rotation angle.
  Bottom: Kurtosis and absolute skewness of the Hillas ellipse.
  }
  \label{fig:proton_hillas_params}
\end{figure}

\paragraph{Hillas size, length, and width}
The Hillas parameters, size,  length, and width, are shown in the top row of figure~\ref{fig:proton_hillas_params}.
For all three parameters, the distributions between the datasets from the MC simulations and the SBDM are generally in good agreement.
Only at the edges of the length distribution deviations are visible, which, however, account for a density less than $5\cdot10^{-3}$ of the images generated by the SBDM.
Comparing the distributions of the WGAN dataset and the MC simulations, significant deviations are visible for all three parameters, with deviations up to $\pm25\,\%$.
Whereas the overall shape of the distributions can be modeled by both the WGAN and the SBDM, the SBDM is significantly superior over the whole phase space.

\paragraph{Radial coordinate, polar coordinate, and rotation angle}
In the middle of figure~\ref{fig:proton_hillas_params}, the distributions of the radial coordinate, the polar coordinate, and the rotation angle, encoding information on the image geometry, are shown.
Both the SBDM and WGAN seem to match the MC simulations very well, with no significant difference in quality.
This indicates that these geometrical parameters, which are also modeled well for $\gamma$-ray images, are easily captured by the generative models.

\begin{figure}[t!]
  \centering
  \includegraphics[scale=0.4]{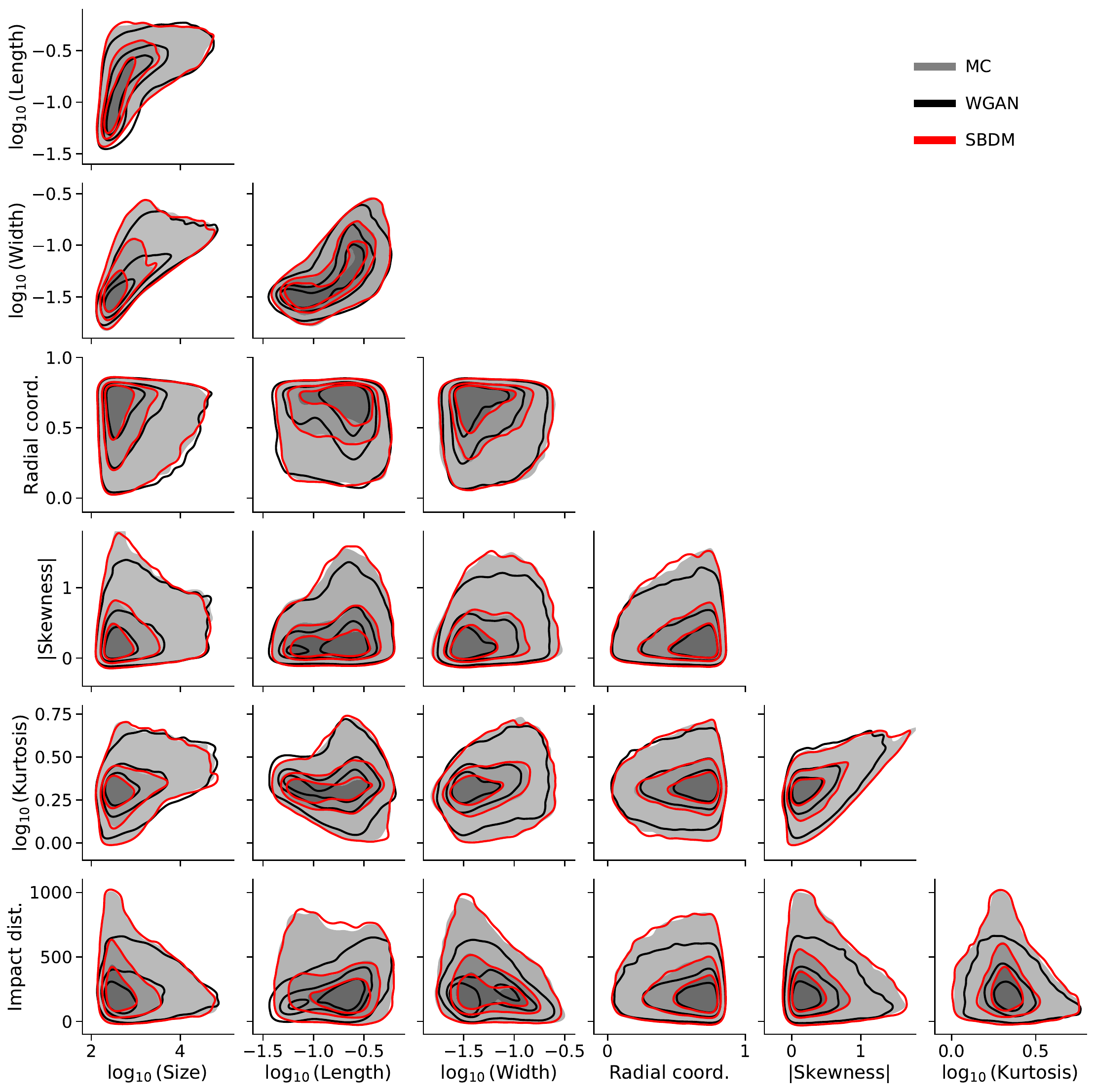}
  \caption{
  Correlations between the air shower impact points and Hillas parameters of the MC (grey), WGAN (black) and SBDM (red) proton dataset.
  The correlations are displayed as kernel density estimation levels with the inner regions corresponding to higher densities.
  } \label{fig:proton_correlations}
\end{figure}

\paragraph{Kurtosis and Skewness}
The distributions for the kurtosis and the skewness of the Hillas ellipse can be found at the bottom of figure~\ref{fig:proton_hillas_params}.
While both models seem to capture the overall shapes of the distributions, distinctive differences between the SBDM and the WGAN are observed.
While the SBDM reproduces the parameter distributions very well, including the distribution tails, the WGAN  struggles to reproduce the tail of the skewness and the bump of the kurtosis distributions, with deviations in the order of $\pm 25\,\%$.\newline

In conclusion, whereas both approaches seem to reproduce the shape of the distributions, the SBDM outperforms the WGAN in terms of quality, highlighted by the ability to capture complex parameter distributions such as skewness and kurtosis, as well as model well the size distributions.
In contrast to $\gamma$-ray differences, which were subtle, generating the more complex proton images seems to reveal limitations of the WGAN approach.

\subsection{Correlations}
\label{sec:proton_correlations}
In figure~\ref{fig:proton_correlations} we show the correlations between the different high-level parameters for the proton dataset, similar to those for the $\gamma$-ray analysis discussed in section~\ref{sec:gamma_correlations}.
In contrast to $\gamma$-ray images, where both approaches seem to reproduce the correlations well, the SBDM and WGAN show different levels of fidelity, with the SBDM demonstrating improvements over the WGAN.
For most cases, distinctive differences between the WGAN and MC data are observed as expected, since discrepancies in the marginals were previously observed.
In particular, correlations between the size, width, kurtosis, skewness, and length parameters are only very qualitatively reproduced by the WGAN, whereas the SBDM achieves a higher level of agreement. 
This demonstrates that the SBDM is superior at generating proton images compared to the WGAN and accurately reproduces even complex physics correlations.
Only low-density regions, indicated by KDE levels with low intensity, show disagreements.

A quantitative analysis of the correlations as described in detail in section~\ref{sec:benchmark} supports these observations as shown in table~\ref{tab:correlation_proton}.
As in the case for the $\gamma$-ray events, the correlations of the SBDM are very close to the optimum, emphasizing that the model can accurately generate proton images and their higher-level features.
In contrast, the WGAN performance is significantly worse and shows a clear degradation in performance compared to the task in generating $\gamma$-ray images, underpinning the qualitative analysis of the visual presentation of the correlations.
This again highlights the challenge of WGAN in producing realistic proton images, particularly with respect to correlations.

\begin{table}[t!]
\centering
\begin{centering}
\begin{tabular}{ c | c | c | c }
\hline
& {Frobenius} & {Correlation matrix} & {Accuracy} \\
& {distance} & {similarity} & {random forest} \\
\hline
Optimal case & 0 & 1 & 0.5 \\
MC vs. WGAN & $0.82 \pm 0.01$ & $0.9154 \pm 0.0028$ & $0.718\pm0.007$ \\
MC vs. SBDM & $0.10\pm0.01$ & $0.9988\pm0.0003$ &$0.522\pm0.011$ \\
\hline
\end{tabular}
\caption{
Quantitative performance measures: Frobenius distance, correlation matrix similarity, and random forest accuracy for the generated proton datasets comparing the WGAN and the SBDM with its similarity to MC. As a reference, the optimal values using indistinguishable datasets are shown in the first row.}
\label{tab:correlation_proton}
\end{centering}
\end{table}

\section{Gamma-hadron separation applied to generated and simulated events}

\begin{figure}[t!]
  \centering
  \includegraphics[scale=0.34]{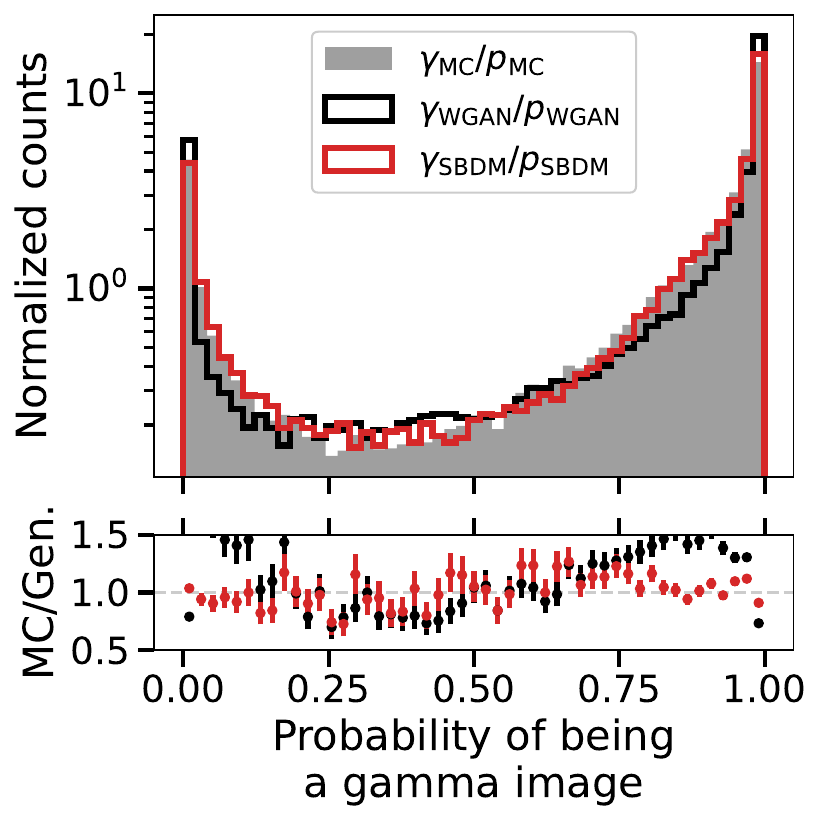}
  \hspace{0.2cm}
  \includegraphics[scale=0.34]{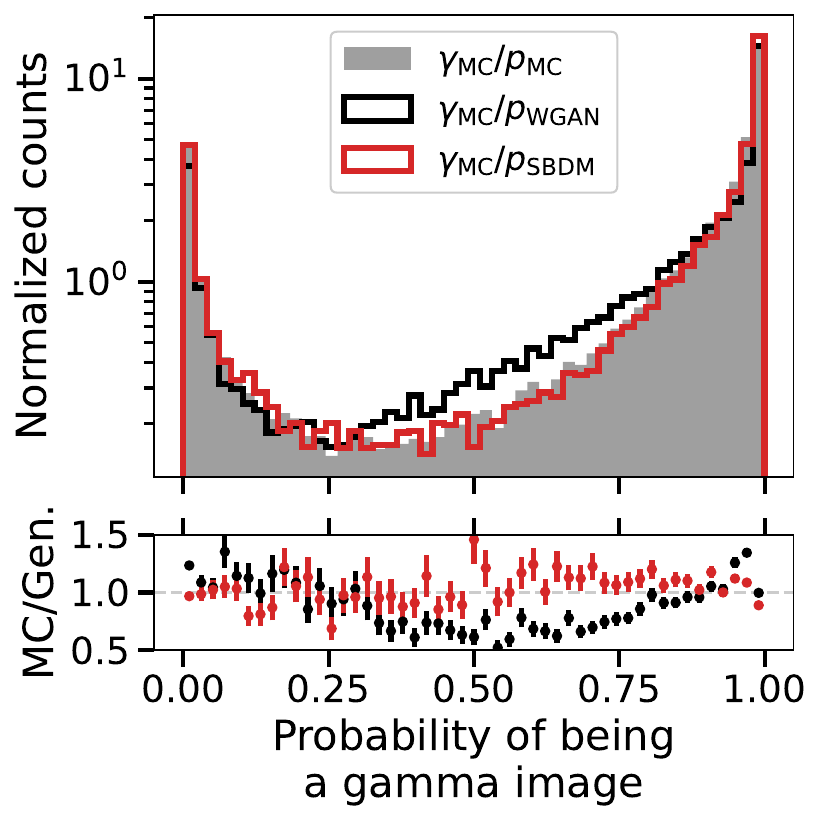}
  \hspace{0.2cm}
  \includegraphics[scale=0.34]{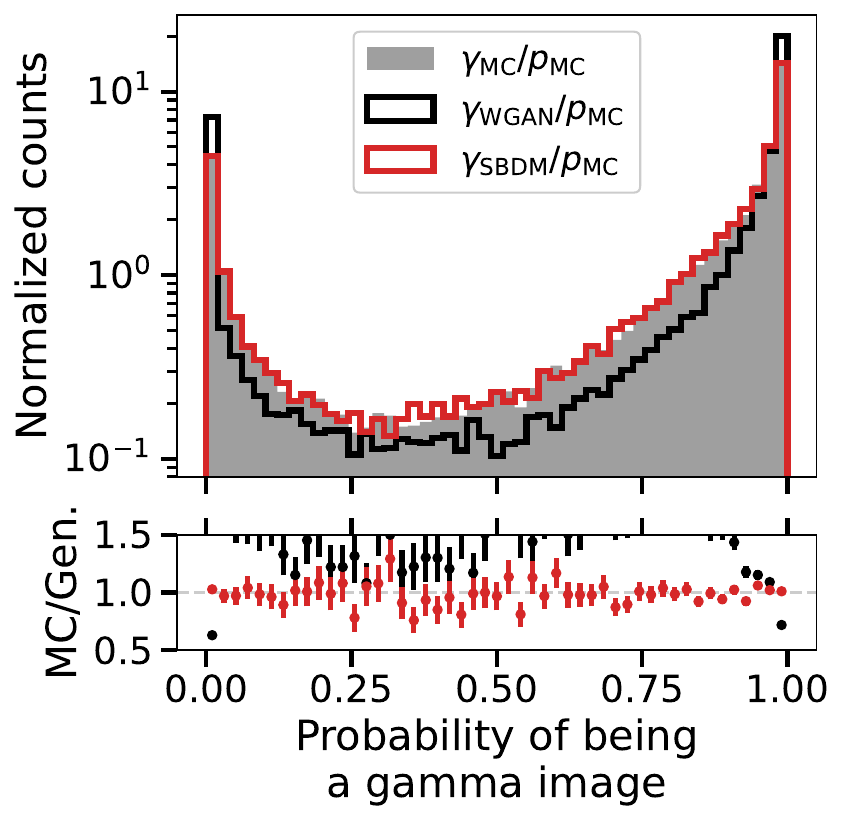}
  \includegraphics[scale=0.33]{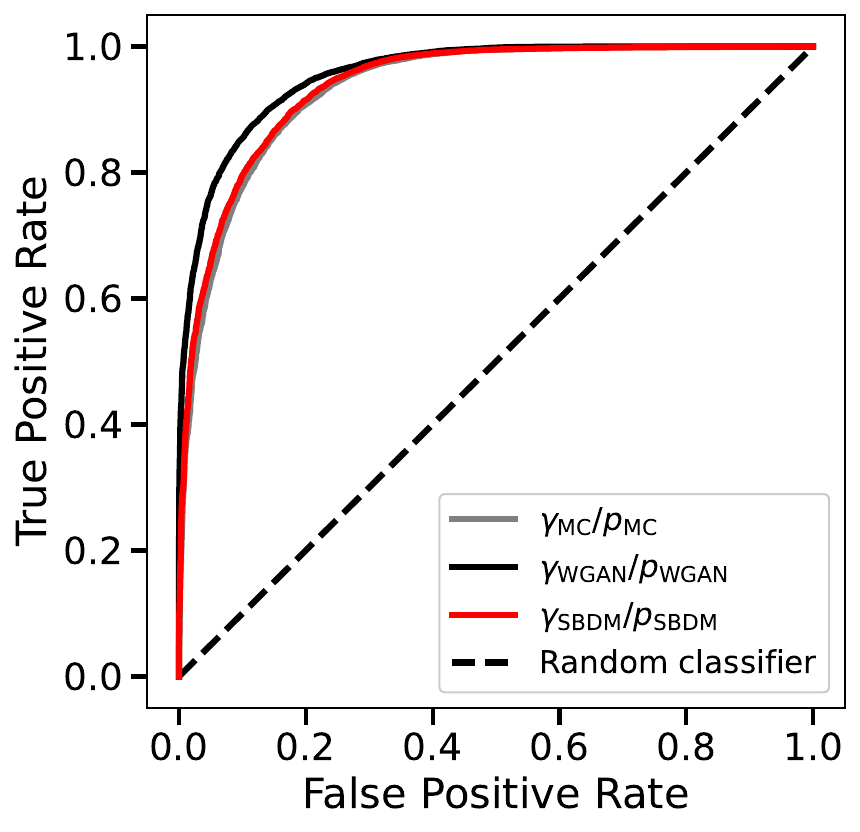}
  \hspace{0.125cm}
  \includegraphics[scale=0.33]{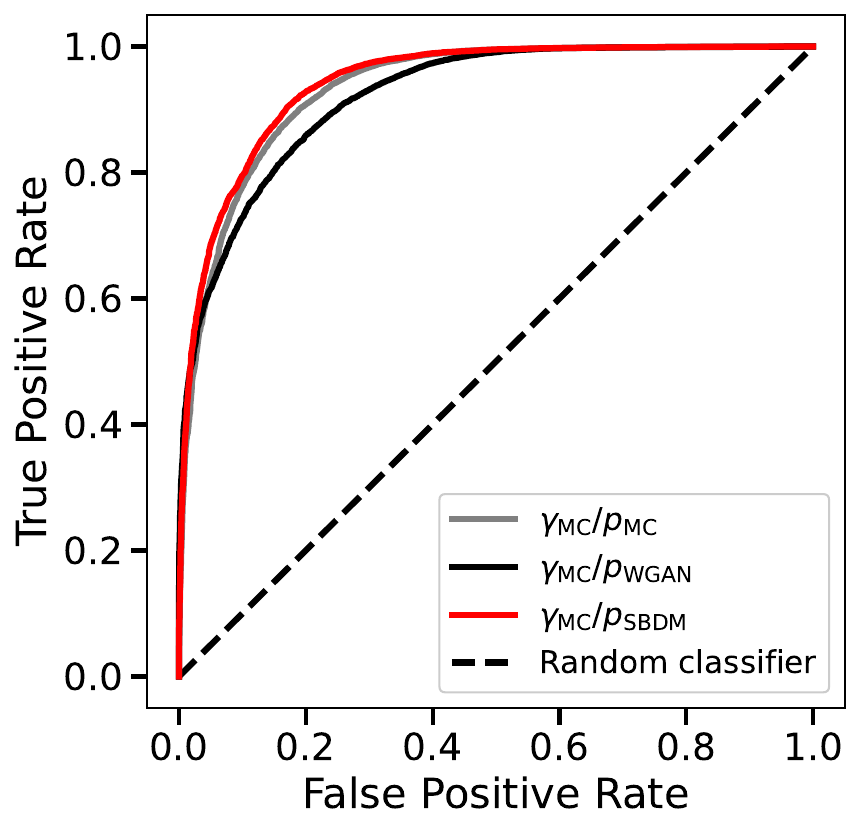}
  \hspace{0.125cm}
  \includegraphics[scale=0.33]{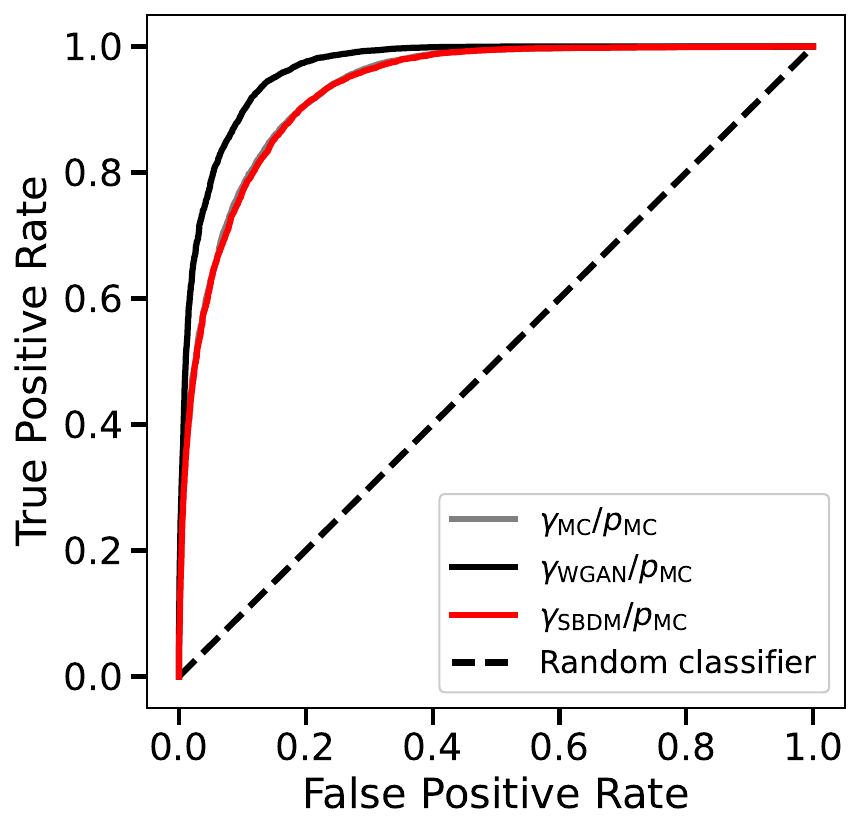}
  \caption{BDT output distributions (top) and ROC curves (bottom) for different combinations of $\gamma$-ray and proton datasets.
  The training and evaluation is carried out on the same type of combination.
  The distributions show the combined results from the $\gamma$-ray and proton evaluations.}
  \label{fig:BDT_results}
\end{figure}

In the previous sections, we evaluated the quality of the generated $\gamma$-ray and proton images using both quantitative and qualitative comparisons.
As the final step of this work, we apply standard $\gamma$-hadron separation, as employed in the H.E.S.S.\ analysis chain, to both the generated and simulated datasets.
This allows us to assess the extent to which current generative models produce images directly usable for downstream physics analyses and quantify their readiness for applications at the analysis-level.
In $\gamma$-ray astronomy, the commonly used method for event classification is Boosted Decision Trees (BDTs)~\cite{OHM2009383, veritas_bdt, cta_design_2011, Becherini_2012, Unbehaun_2025}.
For the BDT training in this work, the number of trees is set to 200, the maximum depth to 6, and the learning rate to 0.3.
The input to the BDT consists of the following parameters: the Hillas width, length, length over $\log_{10}$(size), Kurtosis, absolute value of the Skewness, radial coordinate, and the number of signal pixels, following the standard strategy of state-of-the-art mono analyses~\cite{Unbehaun_2025}.
We refer to the BDT output as \emph{gammaness}, i.e., the likelihood that the event is a $\gamma$-ray.
Model performance is evaluated by  Receiver Operating Characteristic (ROC) curves, showing the True Positive Rate (TPR) over the False Positive Rate (FPR).
While we define the former as the fraction of $\gamma$-rays that are correctly classified, the latter gives information about how many protons are falsely identified as $\gamma$-rays.

We carry out the $\gamma$-hadron separation for seven combinations to examine in detail how realistic the generated datasets are.
First, we quantify how well the generative models represent the differences between these two types of images. Thus, we train and evaluate the BDT on the following combinations: $\gamma_{\text{MC}}/p_{\text{MC}}$ (grey), $\gamma_{\text{WGAN}}/p_{\text{WGAN}}$ (black), $\gamma_{\text{SBDM}}/p_{\text{SBDM}}$ (red).
The corresponding gammaness and ROC curves are shown in the first column of figure~\ref{fig:BDT_results} (left).
While the overall gammaness distributions are similar for all models, the classified output of the SBDM (red histogram) dataset matches the MC dataset (filled grey histogram) significantly better than the WGAN (black histogram).
This is also reflected in the ROC of the SBDM (red), which closely matches the MC (grey), whereas the WGAN (black) curve shows overly strong separation.
This indicates that the WGAN produces unrealistic features in either the $\gamma$-ray or proton images that are not present in the simulation.

Next, we train and evaluate the BDT on combinations that include MC $\gamma$-ray images, as shown in the second column of figure~\ref{fig:BDT_results} (center), to study how well the proton images are generated, as they are classified together with simulated $\gamma$-ray images.
As a reference, we also show the combination of $\gamma_{\text{MC}}/p_{\text{MC}}$ (grey). In both the histogram and the ROC we can see that the evaluation of MC gammas and SBDM protons (red) yields results very similar to the evaluation of the combined MC dataset. On the other hand, clear differences are seen between MC and WGAN (black). The observation that the separation worsens for $\gamma_{\text{MC}}/p_{\text{WGAN}}$, compared to $\gamma_{\text{MC}}/p_{\text{MC}}$ indicates that the WGAN proton images do not feature all the details of the simulated proton showers.
This effect is underscored by the observation that the classification between the WGAN $\gamma$-ray dataset and the MC proton dataset is different.
Here, the separation between $\gamma_{\text{WGAN}}/p_{\text{MC}}$ (black) is less pronounced than for the pure MC reference case (grey), indicating mis-modelling in the WGAN proton images. 
In contrast, the SBDM $\gamma$-ray images (red) and MC proton images (grey) have similar classification scores and ROC curves.

\begin{figure}[t!]
  \centering
  \includegraphics[scale=0.425]{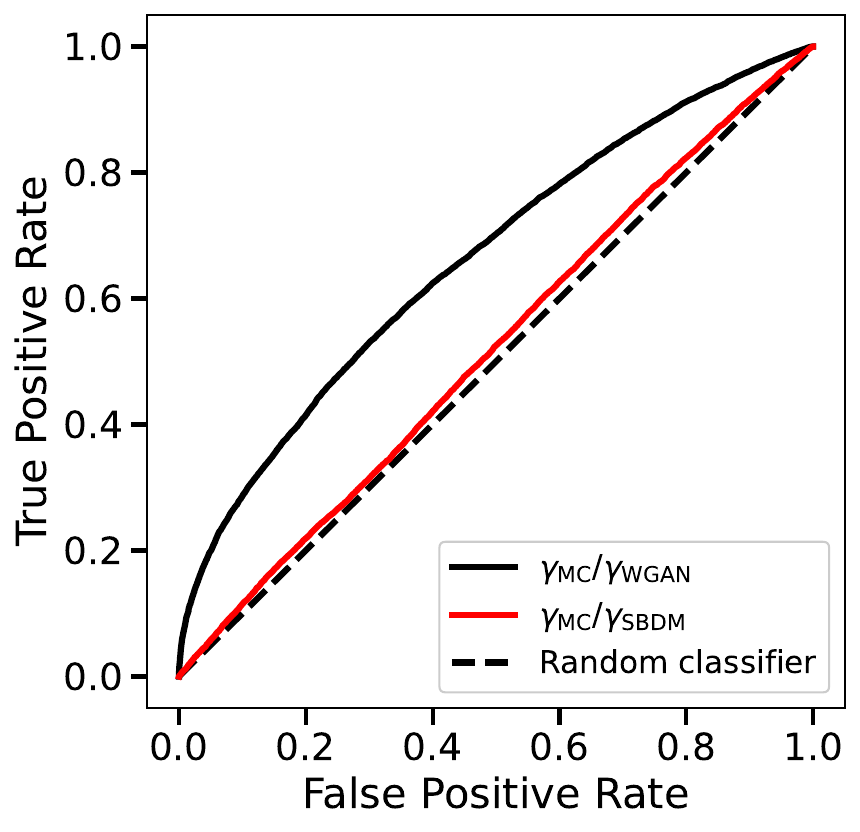}
  \hspace{1.cm}
  \includegraphics[scale=0.425]{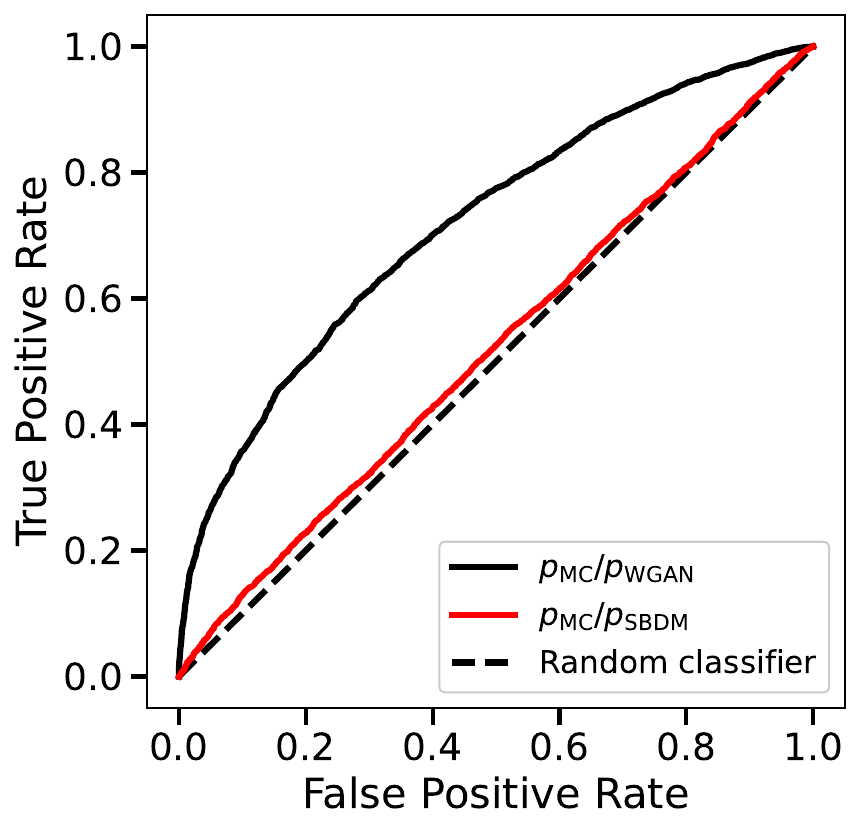}
  \caption{ROC curves for the separation of simulated and generated gamma images (left) and proton images (right) using BDTs.
  The training and evaluation is carried out on the same type of combined dataset.}
  \label{fig:BDT_results_2}
\end{figure}

As the last step of this work, we investigate the separation between generated and simulated images to examine how well the high-level parameters and their correlations are reflected in the generated samples.
In figure~\ref{fig:BDT_results_2}, the ROC curves for different cases are shown.
For both $\gamma_{\text{MC}}/\gamma_{\text{WGAN}}$ (black) and $p_{\text{MC}}/p_{\text{WGAN}}$ (black), the ROC curve deviates significantly from the random classifier (dashed black line) indicating the BDT is able to clearly separate simulated MC and generated WGAN events.
Particularly, the proton images can be easily separated by the BDT, reflecting the challenges of modeling the complex proton signatures using WGANs.
In contrast, the ROC curves for $\gamma_{\text{MC}}/\gamma_{\text{SBDM}}$ (red) and $p_{\text{MC}}/p_{\text{SBDM}}$ (red) are remarkably close to the identity, i.e., the performance of random guessing.
This implies that the BDT cannot separate between the generated SBDM and MC events, making generated samples from the SBDM approach accurate and analysis ready.

This study reveals that while crucial observables are reproduced in both the SBDM and the WGAN models, the level of detail in each approach differs greatly.
While both the proton and $\gamma$-ray images of the WGAN contain features and/or correlations that can be exploited by the BDT, causing performance mismatches of $\gamma$-hadron separation between generated WGAN events and simulated events, the SBDM provides realistic, high-quality data.

\section{Conclusion}

This study presents the first application of diffusion models for generating air-shower images from Imaging Air Cherenkov Telescopes.
Our approach used simulations of the CT5 telescope at H.E.S.S., featuring a modern camera design with close to 2000 pixels, developed for the upcoming CTAO flagship project.
In contrast to previous work that focused on $\gamma$-ray showers, we investigated the generation of both $\gamma$-ray and proton showers with larger and more complex shower-to-shower fluctuations, which are more challenging to model.
Using score-based diffusion models (SBDMs) with a U-Net backbone, we generated $\gamma$-ray and proton showers and compared them against the Wasserstein GAN approach, which showed strong performance in $\gamma$-ray shower generation.
While the SBDM achieves more modest speed-ups relative to the WGAN, reaching factors of approximately 5 on CPUs and up to $5 \times 10^3$ on GPUs compared to the WGAN’s speed-ups of about $2 \times 10^3$ (CPU) to $10^6$ (GPU), the generated samples are substantially better in quality.
Future work will further focus on accelerating the generation times of diffusion models~\cite{progressive_distillation, lyu2022accelerating, zhang2022fast}.

Although the WGAN produces high-quality $\gamma$-ray images, we found that it does not generate proton images of similar quality to MC simulations across both low- and high-level observables.
In contrast, the generated SBDM $\gamma$-ray samples showed improvements over the WGAN for low-level and high-level parameters, as well as reproducing high-level correlations.
In particular, in proton image generation, the SBDM significantly outperformed the WGAN approach, producing proton events with very good agreement with simulations.
At extreme values of the width and length distributions, we observed small disagreements generated by the SBDM, which we plan to focus on in future work.
Finally, we started investigating the analysis readiness of the generated SBDM and WGAN events, i.e., the $\gamma$-ray and proton events, by benchmarking the $\gamma$-hadron separation performance using state-of-the-art techniques.
We found that the SBDM-generated events exhibit little to no deviation from the simulated reference, indicating high fidelity for both $\gamma$-ray and proton images.
In contrast, the WGAN-based approach does not achieve a comparable level of analysis readiness.
Discrepancies are observed for both shower types, reflecting limitations in the generation quality of the WGAN, particularly its reduced ability to reproduce higher-order correlations in the simulated data.
While full analysis-readiness requires further investigations, e.g., the energy or direction reconstruction, the carried out background rejection is already the first significant step towards this final objective.

This work demonstrates that score-based diffusion models can serve as high-fidelity surrogate models for IACT simulations, reproducing not only low and high-level image distributions for both signal and background events but also their internal correlations.
Beyond simply accelerating production, such surrogate models enable rapid re-simulation under changing detector conditions and offer a perspective for end-to-end detector design optimization, unfolding, systematic uncertainty studies, and anomaly detection.
Furthermore, given their differentiable nature, they naturally interface with gradient-based reconstruction and analysis methods, opening the way toward fully differentiable, end-to-end analysis pipelines in astroparticle physics.
Integrating stereoscopic image generation into the generative-modeling framework is a crucial next step.
To date, deep-learning–based approaches proposed by the community have consistently struggled to capture and exploit the stereoscopic nature of IACT images, despite their central role in event reconstruction.
Promising research directions include more complex attention-based architectures that model inter-telescope dependencies, conditional generation based on array geometry and air-shower parameters, as well as hybrid methods that blend generative modeling with physical constraints~\cite{schwefer2024hybridapproacheventreconstruction}.

\section*{Acknowledgments}
We thank the H.E.S.S.\ Collaboration for allowing us to use H.E.S.S.\ simulations for this publication and the H.E.S.S.\ simulation team for running and producing the simulations.
We thank Benedetta Bruno for preparing the simulations and valuable discussions. We also gratefully acknowledge the scientific support and HPC resources provided by the Erlangen National High Performance Computing Center (NHR@FAU) of the Friedrich-Alexander-Universit\"at Erlangen-N\"urnberg (FAU) under the NHR project b129dc.
NHR funding is provided by federal and Bavarian state authorities.
NHR@FAU hardware is partially funded by the German Research Foundation (DFG) – 440719683.
J.\;G. gratefully acknowledges the support of BaCaTeC provided by the Bavarian state authority and the outstanding hospitality of the Machine Learning for Fundamental Physics group at LBNL.
V.\;M. is supported by JST EXPERT-J, Japan Grant Number JPMJEX2509.

\appendix
\section{Pre- and post-processing of data}
\label{appendix}

\paragraph{Generation of zero-value pixels.}

One big challenge for the diffusion models is the generation of pixels with no signal.
A commonly applied solution for this is the addition of noise onto these pixels~\cite{Krause:2021ilc}, which is only used for training and removed after the image generation.
However, for our data, this cannot be done directly, as the distribution of this artificial noise is joint with the distribution of low signal values.
Therefore, we decided to create a gap in the data around zero, in which we can place the noise, such that it is disjoint from the low-signal values.
After the generation of the images, the random noise is then set to zero without affecting any other pixel values.
In our data, we shifted all negative signal values by -1 and all positive values by +1.
This is followed by an addition of Gaussian noise to all pixels using a mean of 0 and standard deviation of 0.2 with a clipping value of $\pm0.9$.
After the generation of new images, all values in [-1, 1] are set to zero, and the negative and positive pixels are shifted back by +1 and -1, respectively.

\paragraph{Conditioning parameters.}

While the diffusion model is able to generate realistic $\gamma$-ray images, some challenges remained for the generation of proton images, which could be explained by their more complex image signal.
Accordingly, apart from the conditioning on the image size of the 4/7 extended cleaning and the $x$- and $y$-coordinate of the air shower impact point, the diffusion model needs more information for the best results.
The new parameters used as additional conditioning are the image size and the number of pixels with signals for various cleaning thresholds.
These tail-cuts cleaning thresholds are 4/7, 5/10, 6/12, 7/14, 9/16 and 10/20.
Thus, the training on proton data uses a total of 15 conditioning parameters, which are predicted with the size model and given to the image model alongside the energy.
In the case of the training on $\gamma$-ray data, we chose to additionally use the image size and number of signal pixels for a 9/16 cleaning as additional conditioning parameters, as we overall saw slight improvements in the results.

\paragraph{Normalizing the input parameters.}

Lastly, the inputs for the two networks of the diffusion model have to be prepared.
First, every image is shifted up by the minimum pixel value of the whole dataset, so that every value is non-negative.
Then all images are divided by their respective size after a shift for the removal of correlation.
For the same reason, the sizes of the images cleaned with 4/7 extended cleaning are divided by the particle energy.
Afterwards, the same functions and transformations are applied to the images, the 4/7 extended sizes and the air shower impact parameters. 
They are scaled to [0, 1] followed by an inwards shift of $10^{-6}$.
Afterwards, a logit transformation and a z-score normalization is carried out.
For the normalization of the six additional sizes, the scaling factor of the initial size is used.
The number of signal pixels are scaled using maximum pixel value of the dataset, as this one stays constant for all cleanings.
In the case of the input energy, the logarithm is taken and the values are shifted to [0, 1].

\paragraph{Input and pre-processing for WGAN.}
The preprocessing of the input parameters generally follows the logarithmic rescaling introduced in ref.~\cite{Elflein_2024}, which is adapted to enable the generation of images with negative values.
To enable the generation of images with negative values, we considered negative values in the training and extended the preprocessing by shifting the values by the lowest value $s_\mathrm{min}$ found in the training dataset before applying the logarithm.
Therefore, we applied to all pixel values $s_i$ the following transformation to obtain the normalized $s_{i,\mathrm{norm}}$ pixel values:
\begin{equation}
    s'_i = \log_{10}(1+s_i - s_\mathrm{min})
\end{equation}

\begin{equation}
    s_{i,\mathrm{norm}} = \frac{s'_i - \mu_\mathrm{50}}{\sigma}
\end{equation}
where, $\sigma$ is the standard deviation and $\mu_\mathrm{50}$ the median calculated over $s'_i$ of the training dataset.

\paragraph{Post-processing of generated images.}

After the comparison of the generation times, the post-processing of the images should be shortly discussed, as it is the last step towards the final IACT images.
It begins with the renormalization process, which includes the removal of the pixel values between $-1$ and $1$ and the shift back of the remaining values in case of the SBDM.
Next, we mask the three mount pixels around the center of the camera, as signal in these pixels would not exist in images from measurements.
Additionally, we have to clip the high values of the generated datasets at roughly 4176\,p.e., which is the saturation value of the MC data.
This is done, since it is very challenging for the models to learn this trend.
Furthermore, a 4/7 cleaning with the extension of four rows is applied to all datasets, as this is currently the only effective way of removing the noise pixels in the WGAN images and having a proper comparison.
Lastly, the generated images with a size smaller than about 140\,p.e. are removed, since this is the smallest size in the simulated images.

\small{
\bibliographystyle{JHEP}
\bibliography{biblio}
}

\end{document}